\pdfoutput=1
\documentclass[onecolumn, apj, iop, numberedappendix, chicago]{emulateapj}
\slugcomment{{To appear in the Astrophysical Journal.}}
\usepackage{amssymb}
\usepackage{amsmath}
\usepackage{cases}
\usepackage{natbib}
\usepackage{subdepth}
\usepackage{tipa}
\usepackage{units}
\usepackage{hyperref}

\allowdisplaybreaks

\newcommand{\delr}[1]{\frac{\partial #1}{\partial r}}
\newcommand{\delth}[1]{\frac{\partial #1}{\partial \theta}}
\newcommand{\delphi}[1]{\frac{\partial #1}{\partial \phi}}
\newcommand{\delrsq}[1]{\frac{\partial^2 #1}{\partial r^2}}
\newcommand{\delthsq}[1]{\frac{\partial^2 #1}{\partial \theta^2}}
\newcommand{\delphisq}[1]{\frac{\partial^2 #1}{\partial \phi^2}}
\newcommand{\delrth}[1]{\frac{\partial^2 #1}{\partial r\partial \theta}}

\newcommand{\delthphi}[1]{\frac{\partial^2 #1}{\partial \theta\partial \phi}}
\newcommand{\xir}{\xi_r}
\newcommand{\xith}{\xi_{\theta}}
\newcommand{\xiphi}{\xi_{\phi}}
\newcommand{\Br}{B_r}

\newcommand{\Bth}{B_{\theta}}

\newcommand{\Bphi}{B_{\phi}}
\newcommand{\Bphih}{\hat{B}_{\phi}}
\newcommand{\sinth}{\sin\theta}
\newcommand{\sinthsq}{\sin^2\theta}
\newcommand{\costh}{\cos\theta}
\newcommand{\cotth}{\cot\theta}
\newcommand{\cotthsq}{\cot^2\theta}
\newcommand{\gsharm}[3]{Y_{#1}^{#2,#3}}
\newcommand{\sharm}[2]{Y_{#1}^{#2}}
\newcommand{\om}[2]{\Omega_{#1}^{#2}}

\newcommand{\D}{\mathrm{d}}
\newcommand{\wtj}[6]{\begin{pmatrix} #1 & #2 & #3 \\ #4 & #5 & #6 \end{pmatrix}}
\newcommand{\p}{'}

\shorttitle{General Matrix Element for Toroidal Magnetic Fields}
\shortauthors{Kiefer et al.}
\begin{document}
	
\title{The Direct Effect of Toroidal Magnetic Fields on Stellar Oscillations:\newline An Analytical Expression for the General Matrix Element}
		
\author{Ren\'e Kiefer}
\affiliation{Kiepenheuer-Institut f\"ur Sonnenphysik, Sch\"oneckstra\ss e 6, 79104 Freiburg, Germany}
	
\author{Ariane Schad}
\affiliation{Kiepenheuer-Institut f\"ur Sonnenphysik, Sch\"oneckstra\ss e 6, 79104 Freiburg, Germany}
\affiliation{Center for Biological Systems Analysis, University of Freiburg, Habsburgerstra\ss e 49, 79104 Freiburg, Germany }

\author{Markus Roth}
\affiliation{Kiepenheuer-Institut f\"ur Sonnenphysik, Sch\"oneckstra\ss e 6, 79104 Freiburg, Germany}

\begin{abstract}
	Where is the solar dynamo located and what is its modus operandi? These are still open questions in solar physics. Helio- and asteroseismology can help answer them by enabling us to study solar and stellar internal structures through global oscillations. The properties of solar and stellar acoustic modes are changing with the level of magnetic activity. 
	However, until now, the inference on subsurface magnetic fields with seismic measures has been very limited.
	The aim of this paper is to develop a formalism to calculate the effect of large-scale toroidal magnetic fields on solar and stellar global oscillation eigenfunctions and eigenfrequencies.
	If the Lorentz force is added to the equilibrium equation of motion, stellar eigenmodes can couple. In quasi-degenerate perturbation theory, this coupling, also known as the direct effect, can be quantified by the general matrix element. We present the analytical expression of the matrix element for a superposition of subsurface zonal toroidal magnetic field configurations.
	The matrix element is important for forward calculations of perturbed solar and stellar eigenfunctions and frequency perturbations. 
	The results presented here will help to ascertain solar and stellar large-scale subsurface magnetic fields, and their geometric configuration, strength, and their change over the course of activity cycles. 
\end{abstract}
	
	

\section{Introduction}\label{sec:intro}
So far, the main benchmarks for dynamo simulations are the correct cycle lengths as well as the properties and distribution of active regions over the simulated cycle \citep[e.g.,][]{Charbonneau2010, Charbonneau2013}. The strength and configuration of large-scale flow fields are important ingredients of modern simulations of solar and stellar flux transport dynamos \cite[e.g.,][]{Choudhuri1995, Dikpati2009, Karak2014}. Seismology has contributed to the refinement of dynamo simulations by mapping the differential rotation in the Sun \citep{Schou1998, Howe2009} and is beginning to do so in stars \citep{Beck2012, Nielsen2015, Hekker2016}. The meridional circulation is another crucial element in these dynamo models, which can be assessed by helioseismic investigations \citep{Giles1997, Braun1998, Haber2002, Schad2013, Zhao2013a}. If it were made possible to measure the location, geometrical configuration, and strength of the magnetic field distribution within the Sun and stars, this would add a benchmark of supreme importance for dynamo simulation.

In the Sun, a large fraction of the total magnetic flux is assumed to be stored in the subsurface toroidal component of the magnetic field \cite[][and references therein]{Charbonneau2010}. The unsigned magnetic flux attributed to active regions amounts to ${\sim}\unit[10^{25}]{Mx}$ over a solar cycle, while the polar cap flux totals to just ${\sim}\unit[10^{22}]{Mx}$ \citep{Charbonneau2010}. As the active regions are thought to originate from deep-seated toroidal flux concentrations \citep[e.g.,][]{Caligari1995, Fan2009}, it is reasonable to concentrate on the toroidal field in a first step, as we will do in this article.

The effect of a magnetic field on stellar acoustic oscillations is twofold: the direct effect, or, as we shall also call it in the following, the coupling of oscillation modes, and the indirect effect.
In this article, we focus on the direct effect. The indirect effect, which is due to additional forces -- in comparison to the equilibrium stellar model without magnetic fields -- perturbs stellar structural quantities, such as sound speed and density. Thus, the cavities where resonant acoustic modes exist are changed, which leads to changes in frequencies and eigenfunctions. The effect of a magnetic field on stellar structure was studied by, e.g., \citet{Mestel1977}, \citet{Mathis2005}, \citet{Duez2008, Duez2010}.

We use an ansatz from quasi-degenerate perturbation theory to calculate the strength of the coupling between stellar oscillation modes \citep{Lavely1992}, which is due to a superposition of large-scale axisymmetric zonal toroidal magnetic fields. The mode coupling leads to frequency shifts, frequency splittings, and distortions of the mode eigenfunctions. The strength of the coupling is quantified by the general matrix element, which is used to construct the supermatrix \citep{Lavely1992}. Solving the eigenvalue problem for the supermatrix yields two essential results. First, the perturbations to the equilibrium frequencies are given by the eigenvalues. Second, the eigenvectors contain the expansion coefficients of the perturbed eigenfunction as components. 

Previous studies with large-scale flows as the source of perturbation have shown that the analysis of perturbed eigenfrequencies and eigenfunctions can be employed to infer the geometrical configuration and strength of that perturbation. \cite{Ritzwoller1991a} developed a formalism to invert global helioseismic data for differential rotation. This was expanded by \cite{Lavely1992} to specifically include global-scale convection and expanded in such a way that general flow fields can be taken as a source of perturbation. With their theory, the perturbation of the mode eigenfrequencies and the perturbed eigenfunctions can be computed. 
In that framework, the effect of the meridional circulation on solar eigenmodes was studied by, e.g., \cite{Roth2008}. Later, \cite{Schad2011} advanced the method by introducing the amplitude ratio of oscillations in the Fourier domain as a helioseismic measurement quantity. With this, \cite{Schad2013a} inferred a multicellular meridional circulation in both depth and latitude by analyzing perturbed solar eigenfunctions.

Using a different perturbation approach, \cite{Gough1990} presented a formalism to calculate the eigenfrequencies of a stellar model with a magnetic field and rotation as perturbations. This formalism included both the direct and indirect effects of the magnetic field. \cite{Antia2000} refined their approach by also including effects from general relativity and second-order effects of rotation. Their main results were forward calculations of helioseismic splitting coefficients. They could limit the strength of a toroidal magnetic field near the Sun's tachocline to $\unit[300]{kG}$.
\cite{Baldner2009} used the theory developed by \cite{Gough1990} and \cite{Antia2000} to probe the solar interior for magnetic field concentrations by comparing the calculated splitting coefficients with the observed values. Their results included two very shallow toroidal field distributions and a poloidal field. However, due to the shallowness of the inferred fields, they could not draw conclusions regarding dynamo models, which mostly rely on deeper seated field concentrations \citep[e.g.,][]{Charbonneau2014}. Recently, \cite{Hanasoge2017} developed a formalism to calculate the sensitivity functions of seismic measurements to the Lorentz stress tensor in the stellar interior.

A seismic investigation of the effect of magnetic fields on the oscillations in the Sun and solar-like stars within the perturbation approach developed by \cite{Lavely1992} has not been done so far. The first step of extending their analytical groundwork to fully include magnetic fields is now presented in this paper. We begin by  deriving the perturbed equation of motion for a non-rotating, non-magnetic star in Section~\ref{sec:equil}. The necessary essentials of quasi-degenerate perturbation theory are reviewed in Section~\ref{sec:perturb}. Section~\ref{sec:gme} is dedicated to the derivation of the general matrix element for a magnetic field, before we give its explicit analytical expression of it in spherical coordinates in Section~\ref{sec:calc}. We discuss our findings and conclude in Section~\ref{sec:concl}. 

\section{The Equilibrium Model}\label{sec:equil}
In this Section, we summarize the derivation of the perturbed equation of motion, because it will help us to trace our steps in later Sections. A detailed treatment of this can be found in, e.g., \cite{Unno1989} and \cite{Aerts2010}.

The equation of motion of a parcel of gas in a non-rotating, static, and spherically symmetric star without a magnetic field reads
\begin{align}
\rho\frac{\D \mathbf{v}}{\D t} = -\boldsymbol{\nabla} p + \rho\mathbf{g}\label{eq:sec1:1},
\end{align}
where $\mathbf{v}$ is the velocity due to displacements from the equilibrium position, $p$ is the pressure, $\mathbf{g}$ is the gravitational acceleration, and $\rho$ is the mass density. For such a star, a standard model can be computed, e.g. for the Sun Model S by \cite{Christensen-Dalsgaard1996}. The model supplies the eigenfunctions and eigenfrequencies of adiabatic oscillations as a solution of the eigenvalue equation, which is briefly derived here. 

The Lagrangian perturbation $\Delta$ of a quantity $Q$ is given by
\begin{align}
\Delta Q = \delta Q + (\boldsymbol{\xi} \cdot \boldsymbol{\nabla})Q,\label{eq:sec1:3}
\end{align}
where $\delta Q$ is the Eulerian variation of $Q$ and $(\boldsymbol{\xi}\cdot\boldsymbol{\nabla})Q$ is the term due to the advection of $Q$ by the displacement $\boldsymbol{\xi}$. The Lagrangian perturbation commutes with the total time derivative:
\begin{align}
\left[D / Dt, \Delta\right] = 0. \label{eq:sec1:4}
\end{align}
Furthermore, the following identifications of Lagrangian perturbations hold, which we need to rewrite the perturbed equation of motion:
\begin{align}
\Delta \mathbf{r} &= \boldsymbol{\xi},\label{eq:sec1:5}\\
\Delta\mathbf{v} &= \frac{ \D \boldsymbol{\xi}}{\D t},\label{eq:sec1:6}
\end{align}
where $\mathbf{r}$ is the position vector.
Here, the material time derivative is equal to the partial time derivative
\begin{align}
\frac{\D}{\D t} = \frac{\partial}{\partial t}\label{eq:sec1:8},
\end{align}
since we consider a star without any velocity fields, such as rotation, convection, or meridional circulation.

We start by dividing Equation~(\ref{eq:sec1:1}) by $\rho$, taking the Lagrangian perturbation $\Delta$ of the whole equation, and multiplying by the unperturbed density $\rho_0$. From now on, unperturbed quantities are labeled with a sub- or superscript $0$. These operations, together with Equations~(\ref{eq:sec1:4})-(\ref{eq:sec1:8}), yield
\begin{align}
\rho_0\frac{\partial^2 \mathbf{\xi}}{\partial t^2}= -\Delta\left(\boldsymbol{\nabla} p\right)+\frac{\Delta\rho}{\rho_0}\boldsymbol{\nabla}p_0 + \rho_0\Delta\mathbf{g}.\label{eq:sec1:2}
\end{align}
A change between the gravitational potential and gravitational acceleration can be done via $-\boldsymbol{\nabla}\Phi = \mathbf{g}$.

By using a modal ansatz for the displacements,
\begin{align}
\boldsymbol{\xi}(\mathbf{r},t) = \boldsymbol{\xi}(\mathbf{r})e^{i\omega t}\label{eq:sec1:10},
\end{align}
and making use of the identification $\rho_0\mathbf{g}_0=\boldsymbol{\nabla}p_0$, the perturbed equation of motion finally reads
\begin{align}
-\rho\omega^2\boldsymbol{\xi} &= -\Delta\left(\boldsymbol{\nabla} p\right)+\mathbf{g}_0\Delta\rho + \rho_0\Delta\mathbf{g}\label{eq:sec1:11}.
\end{align}
We identify the right-hand side with the operator $\mathcal{L}_0(\boldsymbol{\xi}^0)$:
\begin{align}
\mathcal{L}_0(\boldsymbol{\xi}^0) = -\Delta\left(\boldsymbol{\nabla} p\right)+\mathbf{g}_0\Delta\rho + \rho_0\Delta\mathbf{g},\label{eq:sec1:12}
\end{align}
so we have
\begin{align}
-\rho_0\omega^2\boldsymbol{\xi}^0 &= \mathcal{L}_0(\boldsymbol{\xi}^0)\label{eq:sec1:13}.
\end{align}
This equation describes the response of the stellar model to a displacement $\boldsymbol{\xi}$ of a parcel of gas given by the Lagrangian perturbation of Equation~(\ref{eq:sec1:1}). This is the eigenvalue equation we sought to derive. The superscript $0$ on the eigenfunctions is added here to signify that they are eigenfunctions of the equilibrium model. By adding additional forces or velocity fields to the equation of motion, we can probe the response of the model and its eigenfunctions to this perturbation. It may be noted that Equation~(\ref{eq:sec1:12}) is equivalent to Equation~B15 in \cite{Lavely1992} and to Equation~28 in \cite{Lynden-Bell1967}.

In this paper, all calculations are performed in spherical polar coordinates. In the following, $r$ stands for the radial distance from the origin, $\theta$ signifies the colatitude, and $\phi$ is the azimuthal angle. As we are working with a spherically symmetric star, the $r$, $\theta$, and $\phi$ components of the eigenfunctions can be written as
\begin{align}
\boldsymbol{\xi}_{l,m}\left(r,\theta,\phi\right) = 
\begin{pmatrix}
\xi_r\left(r,\theta,\phi\right) \\ \xith\left(r,\theta,\phi\right) \\ \xiphi\left(r,\theta,\phi\right)
\end{pmatrix}
= 
\begin{pmatrix}
\xi^r(r) \gsharm{l}{0}{m}(\theta,\phi)\\
\xi^h(r) \delth{}\gsharm{l}{0}{m}(\theta,\phi)\\
\xi^h(r) \frac{1}{\sinth}\delphi{}\gsharm{l}{0}{m}(\theta,\phi) 
\end{pmatrix}.\label{eq:sec1:14}
\end{align}
Here, $\xi^r(r)$ and $\xi^h(r)$ are the real-valued scalar radial and horizontal displacement amplitudes, respectively. The generalized spherical harmonic functions $\gsharm{l}{N}{m}$ are described in detail in Appendix~\ref{sec:app:gsh}. The set of eigenmodes $\{\boldsymbol{\xi}_k\}$, where $k=\left(l,m,n\right)$, with eigenfrequencies $\omega_k$, gives the solutions to Equation~(\ref{eq:sec1:13}). The integer $l$ is the harmonic degree, $m$ is the azimuthal order, with $|m|\le l$, and $n$ is the radial order of the eigenfunction. As we are considering a non-rotating star, modes of the same radial order and harmonic degree $\left(n,l\right)$ but of different azimuthal order $m$ have the same frequency and form a degenerate multiplet \citep[see, e.g.,][]{Unno1989}: $\omega_{k}=\omega_{nl}$.

The unperturbed eigenfunctions are orthogonal,
\begin{align}
\int_{V}\rho_0\overline{\boldsymbol{\xi}_{k\p}^0}\cdot\boldsymbol{\xi}_{k}^0\D V = N_k\delta_{k\p k}\label{eq:sec1:15}
\end{align}
and normalized with the normalization constant
\begin{align}
N_k = \int_{0}^{R}\rho_0\left[\left(\xi^r_{nl}(r)\right)^2+l\left(l+1\right)\left(\xi^h_{nl}(r)\right)^2\right]r^2\D r.\label{eq:sec1:16}
\end{align}
The integral in Equation~(\ref{eq:sec1:15}) extends over the whole stellar volume and the integral in Equation~(\ref{eq:sec1:16}) extends over the stellar radius.

\section{Essentials of Quasi-degenerate Perturbation Theory}\label{sec:perturb}
We briefly review the required essentials of quasi-degenerate perturbation theory here. More detailed treatments can be found in, e.g., \citet{Lavely1992}, \citet{Roth2002}, or \citet{Schad2013}.
We introduce the following perturbation expansions into Equation~(\ref{eq:sec1:13}):
\begin{subequations}
	\begin{align}
	\mathcal{L}_0 &\rightarrow \mathcal{L}_0 + \epsilon \mathcal{L}_1\label{eq:sec2:1a},\\
	\boldsymbol{\xi}^0 &\rightarrow \tilde{\boldsymbol{\xi}}^0 + \epsilon\tilde{\boldsymbol{\xi}}^1\label{eq:sec2:1b},\\
	\omega^2_{j} &\rightarrow \omega_{\text{ref}}^2+\epsilon\left(\omega_1^2\right)_j\label{eq:sec2:1c}.
	\end{align}
\end{subequations}
They express the effect on the eigenfrequencies, eigenfunctions, and equilibrium equation of motion by the operator $\epsilon \mathcal{L}_1$ that accounts for departures from the equilibrium model. The auxiliary quantity $\epsilon$ signifies a small perturbation and helps to keep track of the order of perturbation. The squared eigenfrequency perturbation in first order is given by $(\omega_1^2)_j$.

The perturbation of the equation of motion leads to a coupling of eigenmodes. The coupling due to, e.g., poloidal flow cells was studied by \cite{Herzberg2016} and that due to a meridional flow in the Sun by \cite{Schad2011}. In the framework of quasi-degenerate perturbation theory, the perturbed eigenfunctions $\tilde{\boldsymbol{\xi}}_j$ to zeroth and first orders, i.e., $\tilde{\boldsymbol{\xi}}_j^0$ and $\tilde{\boldsymbol{\xi}}_j^1$, can be described as a weighted sum over the unperturbed eigenfunctions $\boldsymbol{\xi}_k^0$, for which the modes $k$ are within a coupling set $K$ or $K^{\perp}$, multiplied by a coupling coefficient:
\begin{subequations}
	\begin{align}
	\tilde{\boldsymbol{\xi}}_j^0 &= \sum\limits_{k\in K}{c_{jk}\boldsymbol{\xi}_k^0}\label{eq:sec2:2a},\\
	\tilde{\boldsymbol{\xi}}_j^1 &= \sum\limits_{k\in K^{\perp}}b_{jk}\boldsymbol{\xi}_k^0\label{eq:sec2:2b}.
	\end{align}
\end{subequations}
Which modes are within the coupling set $K$ or its complement $K^{\perp}$ depends on the geometrical configuration of the perturbation, which entails angular momentum selection rules, on the considered modes, as well as on the allowed difference in the frequency of the modes. When, e.g., an axisymmetric meridional poloidal flow is the source of perturbation, only modes of the same azimuthal order can couple due to selection rules \cite[e.g.,][]{Lavely1992, Roth2002}. We will discuss the selection rules for toroidal magnetic fields in Section~\ref{sec:selection}. The coupling coefficients $c_{jk}$ and $b_{jk}$ may be complex valued. The frequency difference of modes within $K$ is restricted by the quasi-degeneracy condition:
\begin{align}
\left|\omega_{\text{ref}}^2-\omega_k^2\right|<\Delta\omega^2\label{eq:sec2:3}.
\end{align}
Here, $\Delta\omega^2$ sets the frequency range of the coupling modes within $K$. The reference frequency $\omega_{\text{ref}}$ is typically chosen to be equal or close to the central frequency of a multiplet $j$. Increasing $\Delta\omega^2$ results in more  modes within the coupling set $K$ and therefore in a more accurate expression for the perturbed quantity. This has to be traded off against the higher computational effort necessary for the computation.

We can focus on the determination of the coefficients $c_{jk}$, which are needed to calculate the perturbed eigenfunction in zeroth order $\tilde{\boldsymbol{\xi}}_j^0$ in Equation~(\ref{eq:sec2:2a}). With the perturbation expansion, Equations~(\ref{eq:sec2:1a})-(\ref{eq:sec2:1c}), the following eigenvalue equation can be found \citep{Lavely1992}:
\begin{align}
\sum_{k\in K}c_{jk}Z_{k\p k} = \sum_{k\in K}c_{jk}\left(\omega_1^2\right)_j\delta_{k\p k} \text{ for }k\p \in K\label{eq:sec2:4}.
\end{align}
Here, it was used that a mode $k$ is either in $K$ or $K^{\perp}$, i.e., the equations for the coefficients $c_{jk}$ and $b_{jk}$ decouple. Now, the sought-for coefficients $c_{jk}$ appear as the elements of the eigenvectors of the matrix $Z$. This matrix is called supermatrix, and its elements are determined from
\begin{align}\label{eq:sec2:5}
Z_{k\p k}&= \begin{cases}
H_{k\p k}-\left(\omega_{\text{ref}}^2-\omega^2_{k}\right)\delta_{k\p k} &\text{for }k\p,k\in K,\\
0 & \text{otherwise.}
\end{cases}
\end{align}
It can be shown that the general matrix element $H_{k\p k}$ is given by
\begin{align}
H_{k\p k} = -\int \overline{\boldsymbol{\xi}}^0_{k\p}\cdot\mathcal{L}_1\left(\boldsymbol{\xi}_k^0\right)\,\D V,\label{eq:sec2:6}
\end{align}
where $\mathcal{L}_1$ is the operator of the perturbing force that causes the coupling, introduced in Equation~(\ref{eq:sec2:1a}).

The eigenvectors and eigenvalues of the supermatrix completely determine the perturbed eigenfunctions and frequencies. The squared frequency correction of a mode $k$ is given by the eigenvalue, which belongs exactly to the eigenvector of $Z$ that holds the expansion coefficients $\left\{c_{jk}\right\}_{k\in K}$ for the construction of the perturbed eigenfunction of the mode.

\section{The General Matrix Element for a Magnetic Field}\label{sec:gme}
As a perturbation to the equilibrium, we add the Lorentz force $\mathbf{F}_L = \frac{1}{c}\mathbf{j}\times\mathbf{B}$, where the current is given by
\begin{align}
\mathbf{j} = \frac{c}{4\pi}\boldsymbol{\nabla}\times\mathbf{B}\label{eq:sec3:2},
\end{align}
to the right-hand side of the equation of motion, Equation~(\ref{eq:sec1:1}). We use the CGS-Gauss system for all calculations. Following the same steps as in Section~\ref{sec:equil} leads to   
\begin{align}
-\rho_0\omega^2\boldsymbol{\xi}^0 &= \mathcal{L}_0(\boldsymbol{\xi}^0)+\mathcal{L}_B(\boldsymbol{\xi}^0)\label{eq:sec3:3},
\end{align}
where we made the identification
\begin{align}
\mathcal{L}_B(\boldsymbol{\xi}^0)= \rho_0\Delta\left(\frac{1}{4\pi\rho}\left(\boldsymbol{\nabla}\times\mathbf{B}\right)\times\mathbf{B}\right)\label{eq:sec3:4}.
\end{align}
This expression is the desired perturbation operator for a magnetic field, which can now be inserted into Equation~(\ref{eq:sec2:6}) with $\mathcal{L}_B(\boldsymbol{\xi}^0)= \mathcal{L}_1(\boldsymbol{\xi}^0)$. Doing so yields the general matrix element for the Lorentz force exerted by a magnetic field of general configuration:
\begin{align}
H_{k\p k} = -\frac{1}{4\pi}\int &\overline{\boldsymbol{\xi}}_{k\p}\cdot \left[\left(\left(\boldsymbol{\nabla}\times\delta\mathbf{B}_k\right)\times\mathbf{B}_0\right) +\left(\left(\boldsymbol{\nabla}\times\mathbf{B}_0\right)\times\delta\mathbf{B}_k\right)\right.\notag\\ &\left.+\left(\boldsymbol{\xi}_k\cdot\boldsymbol{\nabla}\right)\left(\left(\boldsymbol{\nabla}\times\mathbf{B}_0\right)\times\mathbf{B}_0\right)
+\left(\boldsymbol{\nabla}\cdot\boldsymbol{\xi}_k\right)\left(\left(\boldsymbol{\nabla}\times\mathbf{B}_0\right)\times\mathbf{B}_0\right)\right]\,\D V,\label{eq:sec3:6}
\end{align}
where Equation~(\ref{eq:app:sec3:1}) was used. 

The Eulerian perturbation of the magnetic field $\delta\mathbf{B}_k$ can be expressed as 
\begin{align}
\delta\mathbf{B}_k= \boldsymbol{\nabla}\times\left(\boldsymbol{\xi}_k\times\mathbf{B}_0\right).\label{eq:sec3:7}
\end{align}
The subscript $k$ of the eigenmodes signifies that they are solutions to Equation~(\ref{eq:sec1:13}). This is in contrast to $\boldsymbol{\xi}$ in the derivation of identity (\ref{eq:sec3:7}) --- see Appendix~\ref{sec:app:beuler} --- which can be any displacement. Identity~(\ref{eq:sec3:7}) is equivalent to the induction equation of ideal magnetohydrodynamics for small perturbations in the magnetic field and the velocity field.

The unperturbed solar or stellar model has a $(2l+1)$-fold degeneracy in each frequency multiplet $(n,l)$. This degeneracy can easily be lifted by introducing a differential rotation profile as a perturbation to the equilibrium model, and then applying the theory of \cite{Lavely1992} as discussed in, e.g., \cite{Roth2002}.

\subsection{Modeling the Toroidal Magnetic Field}
So far, $\mathbf{B}$ has been a magnetic field of any configuration. We now restrict $\mathbf{B}$ to a superposition of purely toroidal fields without a $\phi$ dependence, which is represented in generalized spherical harmonics as
\begin{align}
\mathbf{B} = \mathbf{B}_{\text{tor}} = \sum_{s}\mathbf{B}_{s} = \sum_{s}-a_{s}(r) \delth{} \gsharm{s}{0}{0}(\theta, \phi)\mathbf{e}_\phi\label{eq:sec3:8}
\end{align}
Here, $a_{s}(r)$ are the radial profiles of the toroidal field components and $\gsharm{s}{0}{0}(\theta, \phi)$ are the generalized spherical harmonic functions. Some useful properties of these functions are collected in Appendix~\ref{sec:app:gsh}. The index range of $s$ extends over those even positive integers, which are contributing to the magnetic field of the model. The restriction to even integers ensures opposite polarities in the northern and southern hemispheres.

\subsection{The Matrix Element of a Toroidal Magnetic Field}
With the specialization to a superposition of toroidal magnetic field components (Equation~(\ref{eq:sec3:8})), the general matrix element from Equation~(\ref{eq:sec3:6}) is now given by
\begin{align}
H_{k\p k} = -&\frac{1}{4\pi}\int \overline{\boldsymbol{\xi}}_{k\p}\cdot \left[\left(\left(\boldsymbol{\nabla}\times\sum_{s}\delta\mathbf{B}_{k,s}\right)\times\sum_{s\p}\mathbf{B}_{s\p}\right)+\left(\left(\boldsymbol{\nabla}\times\sum_{s}\mathbf{B}_s\right)\times\sum_{s\p}\delta\mathbf{B}_{k,s\p}\right)\right.\notag\\
&\left.+\left(\boldsymbol{\xi}_k\cdot\boldsymbol{\nabla}\right)\left(\left(\boldsymbol{\nabla}\times\sum_{s}\mathbf{B}_s\right)\times\sum_{s\p}\mathbf{B}_{s\p}\right)
+\left(\boldsymbol{\nabla}\cdot\boldsymbol{\xi}_k\right)\left(\left(\boldsymbol{\nabla}\times\sum_{s}\mathbf{B}_{s}\right)\times\sum_{s\p}\mathbf{B}_{s\p}\right)\right]\,\D V,\label{eq:sec3:9}
\end{align}
where $s$ and $s\p$ extend over the same index range. Due to the distributivity of the vector product and the dot product as well as due to the sum rule for the integral, this simplifies to
\begin{align}
H_{k\p k} = -\frac{1}{4\pi}\sum_{s,s\p}\int \overline{\boldsymbol{\xi}}_{k\p} \cdot &\left[  \left(\left(\boldsymbol{\nabla}\times\left(\boldsymbol{\nabla}\times\left(\boldsymbol{\xi}_k\times\mathbf{B}_{s}\right)\right)\right)\times\mathbf{B}_{s\p}\right)+\left(\left(\boldsymbol{\nabla}\times\mathbf{B}_s\right)\times\left(\boldsymbol{\nabla}\times\left(\boldsymbol{\xi}_k\times\mathbf{B}_{s\p}\right)\right)\right)\right.\notag\\
&\left.+\left(\boldsymbol{\xi}_k\cdot\boldsymbol{\nabla}\right)\left(\left(\boldsymbol{\nabla}\times\mathbf{B}_s\right)\times\mathbf{B}_{s\p}\right)+\left(\boldsymbol{\nabla}\cdot\boldsymbol{\xi}_k\right)\left(\left(\boldsymbol{\nabla}\times\mathbf{B}_s\right)\times\mathbf{B}_{s\p}\right)\right]\,\D V,\label{eq:sec3:10}
\end{align}
where we applied Equation~(\ref{eq:sec3:7}). We will now calculate the extended form of this equation in spherical polar coordinates.

\section{The Analytical Form of the General Matrix Element}\label{sec:calc}
To simplify notation and avoid indices, we consider one summand from Equation~(\ref{eq:sec3:10}) for fixed $s,s\p$. We omit the indices $s,s\p$ and use
\begin{align}
\Bphi &= a(r)\delth{}\gsharm{s}{0}{0}\left(\theta,\phi\right),\label{eq:sec4:1}\\
\Bphih &= \hat{a}(r)\delth{}\gsharm{s\p}{0}{0}\left(\theta,\phi\right),\label{eq:sec4:2}
\end{align}
where $a_s = a$ and $a_{s\p} = \hat{a}$. We use the same notation for the eigenmodes, which appear twice in Equation~(\ref{eq:sec3:10}), once due to the Lagrangian perturbation of the Lorentz force and once for the eigenmode that is multiplied from the left in Equation~(\ref{eq:sec3:10}). The eigenmode that is multiplied from the left is assigned the hat, i.e., its radial component is written as $\hat{\xi}_r = \hat{\xi}^r(r) \gsharm{l\p}{0}{m\p}(\theta,\phi)$.

We line out the necessary calculations in Appendix~\ref{sec:app:calc}. As a result, we find the following expression for the general matrix element given by Equation~(\ref{eq:sec3:10}):
\begin{align}
H_{k\p k}= -\frac{1}{4\pi}\int&\left[
\left(2\hat{a}{\hat{\xi}^r}\delr{a}\delr{\xi^r}+\frac{2a\hat{a}{\hat{\xi}^r}}{r}\delr{\xi^r} +a\hat{a}{\hat{\xi}^r}\delrsq{\xi^r}\right)\mathcal{T}_1
+\left(\frac{\hat{a}{\hat{\xi}^r}\xi^h}{r}\delr{a} +\frac{a\hat{a}{\hat{\xi}^r}}{r}\delr{\xi^h}\right)\mathcal{T}_2\right.\notag\\
&+\left(\frac{a\hat{a}{\hat{\xi}^h}\xi^h }{r^2}+\frac{\hat{a}{\hat{\xi}^h}\xi^h}{r}\delr{a}+\frac{a\hat{a}{\hat{\xi}^r}\xi^r}{r^2}-\frac{a\hat{a}{\hat{\xi}^h}\xi^r}{r^2}-\frac{a\hat{a}{\hat{\xi}^r}\xi^h}{r^2}-\frac{\hat{a}{\hat{\xi}^h}\xi^r}{r}\delr{a}-\frac{\hat{a}{\hat{\xi}^r}\xi^h}{r}\delr{a}\right)\mathcal{T}_{3}\notag\\
&+\frac{a\hat{a}{\hat{\xi}^r}}{r}\delr{\xi^h}\mathcal{T}_4+\left(\frac{a\hat{a}{\hat{\xi}^h}}{r}\delr{\xi^r} +\frac{a\hat{a}{\hat{\xi}^h}\xi^r }{r^2} \right)\mathcal{T}_5-\frac{a\hat{a}{\hat{\xi}^h}\xi^h}{r^2}\mathcal{T}_{6}+\frac{2a\hat{a}{\hat{\xi}^h}\xi^h }{r^2}\mathcal{T}_7\notag\\
&+\left(\frac{a\hat{a}{\hat{\xi}^h}\xi^r }{r^2}+\frac{a\hat{a}{\hat{\xi}^h}}{r}\delr{\xi^r}\right)\mathcal{T}_8+\frac{a\hat{a}{\hat{\xi}^h}\xi^h }{r^2}\mathcal{T}_9+\left(\frac{a\hat{a}{\hat{\xi}^h}\xi^r }{r^2}+\frac{a\hat{a}{\hat{\xi}^h}}{r} \delr{\xi^r}\right.\notag\\
&\left.+\frac{\hat{a}{\hat{\xi}^h}\xi^r}{r}\delr{a}-\frac{\hat{a}{\hat{\xi}^h}\xi^h }{r}\delr{a}\right)\mathcal{T}_{10}+\frac{a\hat{a}{\hat{\xi}^h}\xi^h}{r^2}\mathcal{T}_{11}+\frac{a\hat{a}{\hat{\xi}^h}\xi^h}{r^2}\mathcal{T}_{12}+\frac{a\hat{a}{\hat{\xi}^h}\xi^h}{r^2}\mathcal{T}_{17}+\frac{a\hat{a}{\hat{\xi}^h}\xi^h}{r^2}\mathcal{T}_{18}\notag\\
&-\frac{\hat{a}{\hat{\xi}^r}\xi^h}{r}\delr{a}\mathcal{T}_{20}\left.-\frac{a \hat{a}{\hat{\xi}^h}\xi^h}{r^2}\mathcal{T}_{22}-\frac{a\hat{a}{\hat{\xi}^h}\xi^h}{r^2}\mathcal{T}_{23}-\frac{a\hat{a}{\hat{\xi}^h}\xi^h}{r^2}\mathcal{T}_{24}-\frac{a\hat{a}{\hat{\xi}^h}\xi^h}{r^2}\mathcal{T}_{25}\right]\,\D V\label{eq:sec4:3}
\end{align}
The factors with a radial dependence have been separated from those factors with an angular dependence. The angular parts are contained in the symbols $\mathcal{T}_i$ with $i=1,\dots,25$. They are listed in Appendix~\ref{sec:app:kernels}. The integral in Equation~(\ref{eq:sec4:3}) extends over the whole stellar volume. Integration over the solid angle can be carried out by utilizing Wigner $3j$ symbols, as demonstrated in Appendix~(\ref{sec:app:integral}). The properties of the Wigner $3j$ symbols (see Appendix~(\ref{sec:app:wtj})), can be used to reduce the number of terms in the angular kernels~(\ref{app:eq:kernel:deriv:1})-(\ref{app:eq:kernel:deriv:25}), which leads to Equations~(\ref{app:eq:kernel:fin:1})-(\ref{app:eq:kernel:fin:25}). The summation over the components of a superposition of toroidal magnetic fields as in Equation~(\ref{eq:sec3:7}) can now easily be applied again by assigning the indices $s,s\p$ to the two magnetic fields as defined in Equations~(\ref{eq:sec4:1})-(\ref{eq:sec4:2}).

\subsection{Selection Rules}\label{sec:selection}
From the properties of the integral over four generalized spherical harmonics, given by Equation~(\ref{app:eq11}), we find the following selection rules for mode coupling: two modes can only couple if their harmonic degrees satisfy
\begin{align}\label{eq:sec4:4}
|l-l\p| \le s+s\p.
\end{align}
This follows from property (\ref{app:eq:wtj5.3}) of the Wigner $3j$ symbol.
The azimuthal order of the coupling modes must satisfy
\begin{align}\label{eq:sec4:5}
m=m\p,
\end{align}
which is a consequence of the property (\ref{app:eq:wtj5.1}) of the Wigner $3j$ symbol and the fact that the azimuthal order was set to zero for the toroidal magnetic field. 

\section{Discussion and Conclusion}\label{sec:concl}
Magnetic fields in stars lead to the coupling of stellar oscillation eigenmodes. This coupling perturbs the mode frequencies and eigenfunctions of these acoustic oscillations. With an ansatz from quasi-degenerate perturbation theory, in which the magnetic field is treated as a small perturbation to the equilibrium stellar model, we presented a detailed derivation of the general matrix element. In this work, we focused on the direct effect caused by a superposition of toroidal magnetic fields. The direct effect describes the mode coupling due to the Lorentz force, whereas the indirect effect, which we did not consider in this article, is due to the perturbation of the mode cavities by the Lorentz force. Both the direct and indirect effects have to be accounted for, if a full description of the perturbed mode frequencies and eigenfunctions is to be reached. The combination of both effects will be considered in an upcoming study.

The analytical expression of the general matrix element was obtained with the use of generalized spherical harmonics and the calculation of the angular integral over their product. The resulting general matrix element, presented in Equation~(\ref{eq:sec4:3}), can be used to carry out forward calculations of the direct effect of arbitrary subsurface zonal toroidal field configurations on the stellar eigenmodes. From the general matrix elements, the supermatrix can be constructed. Its eigenvalues and eigenvectors hold the information on the mode frequency perturbations and the perturbed eigenfunctions, respectively. 

From an inspection of the general matrix element and the selection rules for coupling of angular momenta, we find that toroidal magnetic fields couple only modes that are of equal azimuthal order (Equation~(\ref{eq:sec4:5})). Also, only modes whose harmonic degrees  maximally differ by the sum of the harmonic degrees of the toroidal magnetic field configuration can couple (Equation~(\ref{eq:sec4:4})). It is noteworthy that the general matrix element and hence the supermatrix $Z$ is not Hermitian. This is in stark contrast to the supermatrix for toroidal flows, e.g., rotation \citep{Lavely1992}, or poloidal flows as the meridional circulation \citep{Roth2008}, where $Z$ is Hermitian. The Hermiticity of the supermatrix for flows ensures that its eigenvalues and consequently the frequency perturbations are purely real. This need not be the case if the perturbation is a toroidal magnetic field. The non-Hermiticity of the supermatrix, or more precisely its asymmetry as it is completely real-valued in the case of magnetic fields, allows complex-valued eigenvalues and thus complex frequency perturbations. The asymmetry can be easily seen in Equation~(\ref{eq:sec4:3}), considering, for example, the radially dependent part of the first term, which includes the angular factor $\mathcal{T}_1$. Clearly, the terms are not symmetrical under the exchange of the modes $k\p$ and $k$, which renders the general matrix element asymmetrical. A complex-valued frequency perturbation is tantamount to a frequency shift and an additional damping factor for the affected mode. 

The presented matrix element can be used to study the direct effect of toroidal field configurations motivated by the simulation of solar and stellar dynamos, e.g., by \cite{Miesch2016}, in forward calculations for their effect on the solar and stellar eigenmodes. The resulting frequency and damping perturbations can then be compared with observational data for frequency shifts \citep[e.g.,][]{Woodard1985, Libbrecht1990, Jimenez-Reyes1998, Broomhall2017} and mode damping rates \citep[e.g.,][]{Jefferies1990, Komm2000a, Broomhall2015} over the solar cycle as well as the frequency shifts caused by changes in stellar magnetic activity \citep{Garcia2010, Salabert2016a, Kiefer2017}. This will provide a novel tool to calibrate and test dynamo simulations and may even help magnetic field concentrations to be locate within the solar and stellar interiors. With the formalism presented here, it will also be possible to produce artificial splitting coefficients for toroidal fields. These can then also be compared to observed splitting coefficients, which may additionally help in locating subsurface magnetic field structures.

It has been shown in previous studies that the analysis of perturbed eigenfunctions can be used to infer the profile of the solar meridional flow \citep{Schad2013, Schad2013a}. The results from the present work are a first step toward such an inversion with the aim to infer the solar magnetic field from perturbed eigenfunctions analogous to the work by \cite{Schad2013a}. In a next step, also the indirect effect of the magnetic field also has to be accounted for, which is a work in progress. 

In contrast to earlier studies, which aimed to determine the effect of subsurface magnetic fields in the Sun on global oscillations \citep[e.g.,][]{Gough1990, Antia2000, Baldner2009}, we present an analytical expression for the direct effect. In this way, it can readily be investigated how different modes are affected by the input magnetic field depending on the amplitudes of their radial and horizontal displacements, $\xi^r$ and $\xi^h$; their harmonic degree; frequency; and other parameters. Modes of different frequencies and harmonic degrees probe different depths of the star \citep[e.g.,][]{Aerts2010}. Depending on this lower turning point, modes are variably susceptible to perturbations by a magnetic field of a certain location and configuration. In an upcoming study, we will exploit these properties of the presented formalism by inputting different magnetic field configurations and learn how different modes are affected by different magnetic fields.

We note that the analytical expression of the general matrix element for the direct effect of a superposition of poloidal fields or general magnetic field configuration will lead to even more extensive matrix elements than the one presented here. Nonetheless, this will be a worthwhile effort and should be carried out in the future.

With the result of this paper, we contribute to a novel approach to measure subsurface magnetic field concentrations in the Sun and stars. The systematic forward modeling of frequency shifts, frequency splittings, and the distortion of mode eigenfunctions will help us to learn about the location of solar and stellar dynamos and the determination of their mode of operation.

\acknowledgments
The authors wish to thank the anonymous referee and our institute's internal referee Oskar Steiner for taking the time to review this article.

The research leading to these results received funding from the European Research Council under the European Union’s Seventh Framework Program (FP/2007-2013)/ERC Grant Agreement No. 307117.

	
\appendix

\section{The Eulerian Perturbation of the Magnetic Field}\label{sec:app:beuler}
In ideal magnetohydrodynamics, the induction equation can be written as
\begin{align}
\frac{\partial \mathbf{B}}{\partial t} = \boldsymbol{\nabla} \times \left(\mathbf{u}\times\mathbf{B}\right),\label{eq:app:sec1:1}
\end{align}
where $\mathbf{B}$ is the magnetic field and $\mathbf{u}$ is the velocity field.
We introduce the perturbations
\begin{align}
\mathbf{B} \rightarrow \mathbf{B}_0(\mathbf{r})+\delta\mathbf{B}(\mathbf{r},t),\label{eq:app:sec1:2}\\
\mathbf{u} \rightarrow \mathbf{u}_0(\mathbf{r}) + \mathbf{v}(\mathbf{r},t),\label{eq:app:sec1:3}
\end{align}
where $\delta\mathbf{B}$ and $\mathbf{v}$ are the Eulerian perturbation to the magnetic field and the velocity field, respectively.
The unperturbed magnetic field $\mathbf{B}_0(\mathbf{r})$ is assumed to be static. We neglect large-scale velocity fields here, as we are focusing on the influence of the magnetic field. Hence, the background velocity field $\mathbf{u}_0(\mathbf{r})$ is set to zero. The Eulerian perturbation of the velocity field $\mathbf{v}(\mathbf{r},t)$ is described as a temporal change in the displacement $\boldsymbol{\xi}$:
\begin{align}
\mathbf{v} = \frac{\partial \boldsymbol{\xi}}{\partial t}\label{eq:app:sec1:4}.
\end{align}
Applying Equations~(\ref{eq:app:sec1:2})-(\ref{eq:app:sec1:4}) to Equation~(\ref{eq:app:sec1:1}), retaining only terms that are linear in the perturbations, and integrating over time yield the Eulerian perturbation of the magnetic field $\delta\mathbf{B}$:
\begin{align}
\delta\mathbf{B}\left(\mathbf{r},t\right)= \boldsymbol{\nabla}\times\left(\boldsymbol{\xi}\left(\mathbf{r},t\right)\times\mathbf{B}_0\left(\mathbf{r}\right)\right)\label{eq:app:sec1:5}.
\end{align} 

\section{Mathematical Supplements}\label{sec:app:math}
In this appendix, we list the vector identities and formulae that were used in the calculation of the general matrix element in Section~\ref{sec:calc} and Appendix~\ref{sec:app:calc}, and help the traceability of our derivations. All calculations are done in spherical polar coordinates. In the following, $r$ denotes the radial distance from the origin, $\theta$ signifies the colatitude, and $\phi$ is the azimuthal angle.

Let $\mathbf{A}\left(r,\theta,\phi\right)$ and $\mathbf{B}\left(r,\theta,\phi\right)$ denote general vector fields. A useful identity for the curl of the cross-product of two vector fields $\mathbf{A}$ and $\mathbf{B}$ is given by
\begin{align}
\boldsymbol{\nabla}\times\left(\mathbf{A}\times\mathbf{B}\right) = \mathbf{A}\left(\boldsymbol{\nabla}\cdot\mathbf{B}\right)-\mathbf{B}\left(\boldsymbol{\nabla}\cdot\mathbf{A}\right)+\left(\mathbf{B}\cdot\boldsymbol{\nabla}\right)\mathbf{A}-\left(\mathbf{A}\cdot\boldsymbol{\nabla}\right)\mathbf{B}\label{eq:app:sec2:1}.
\end{align}

The divergence of a vector field $\mathbf{A}$ is
\begin{align}
\boldsymbol{\nabla}\cdot\mathbf{A} = \frac{1}{r^2}\delr{\left(r^2A_r\right)}+\frac{1}{r\sinth}\delth{}\left(A_{\theta}\sinth\right)+\frac{1}{r\sinth}\delphi{A_{\phi}}\label{eq:app:sec2:2},
\end{align}
and its curl is
\begin{align}
\boldsymbol{\nabla}\times\mathbf{A} =& 
\frac{1}{r\sinth}\left(\delth{}\left(A_{\phi}\sinth\right)-\delphi{A_{\theta}}\right)\boldsymbol{e}_r\notag\\
&+\frac{1}{r}\left(\frac{1}{\sinth}\delphi{A_r}-\delr{}\left(rA_{\phi}\right)\right)\boldsymbol{e}_{\theta}\notag\\
&+\frac{1}{r}\left(\delr{}\left(rA_{\theta}\right)-\delth{A_{r}}\right)\boldsymbol{e}_{\phi}\label{eq:app:sec2:3}.
\end{align}

The material derivative of the vector fields $\mathbf{A}$ and $\mathbf{B}$ is given by
\begin{align}
\left(\mathbf{A}\cdot\boldsymbol{\nabla}\right)\mathbf{B}=&\left(A_r\delr{B_r}+\frac{A_{\theta}}{r}\delth{B_r}+\frac{A_{\phi}}{r\sinth}\delphi{B_r}-\frac{A_{\theta}B_{\theta}+A_{\phi}B_{\phi}}{r}\right)\boldsymbol{e}_r\notag\\
&+\left(A_r\delr{\Bth}+\frac{A_{\theta}}{r}\delth{\Bth}+\frac{A_{\phi}}{r\sinth}\delphi{\Bth}+\frac{A_{\theta}\Br}{r}-\frac{A_{\phi}\Bphi\cotth}{r}\right)\boldsymbol{e}_{\theta}\notag\\
&+\left(A_r\delr{\Bphi}+\frac{A_{\theta}}{r}\delth{\Bphi}+\frac{A_{\phi}}{r\sinth}\delphi{\Bphi}+\frac{A_\phi\Br}{r}+\frac{A_{\phi}\Bth\cotth}{r}\right)\boldsymbol{e}_{\phi}\label{eq:app:sec2:4}.
\end{align}

\section{Calculation of the General Matrix Element}\label{sec:app:calc}
In this appendix, we derive the explicit expression for the general matrix element in spherical coordinates.
We start from the Lagrangian perturbation of the Lorentz force:
\begin{align}
&\rho_0\Delta\left(\frac{1}{c\rho}\mathbf{j}\times\mathbf{B}\right)\notag\\
=& \frac{1}{4\pi}\left(\left(\left(\boldsymbol{\nabla}\times\delta\mathbf{B}\right)\times\mathbf{B}_0\right) +\left(\left(\boldsymbol{\nabla}\times\mathbf{B}_0\right)\times\delta\mathbf{B}\right)+\left(\boldsymbol{\xi}\cdot\boldsymbol{\nabla}\right)\left(\left(\boldsymbol{\nabla}\times\mathbf{B}_0\right)\times\mathbf{B}_0\right)	+\left(\boldsymbol{\nabla}\cdot\boldsymbol{\xi}\right)\left(\left(\boldsymbol{\nabla}\times\mathbf{B}_0\right)\times\mathbf{B}_0\right)\right)\label{eq:app:sec3:1},
\end{align}
where Equations~(\ref{eq:sec1:3}) and (\ref{eq:sec3:2}), in which the Lagrangian perturbation $\Delta$ and the electric current $\mathbf{j}$ are defined, have been used, and Equation~(\ref{eq:app:sec1:5}) for the Eulerian perturbation of the magnetic field $\delta\mathbf{B}$ holds. Before applying the Lagrangian perturbation to the Lorentz force, we divided by $\rho$ only to multiply by $\rho_0$ afterwards. Due to the Lagrangian perturbation of $1/\rho$, a term with a factor of $-\Delta\rho/\rho_0$ appears. This term was rewritten by virtue of the Lagrangian perturbation of the density $\Delta\rho = -\rho_0\boldsymbol{\nabla}\cdot\boldsymbol{\xi}$, which can be derived by linearizing the perturbed equation of mass conservation. Furthermore, $c$ is the speed of light and $\boldsymbol{\xi}$ is an eigenfunction. 

We restrict the magnetic field to be of toroidal shape but do not yet specify the exact configuration:
\begin{align}\label{eq:app:sec3:2}
\mathbf{B}=\mathbf{B}_{\text{tor}} =  \begin{pmatrix}
0\\0\\B_{\phi}\left(r,\theta,0\right)
\end{pmatrix}.
\end{align}
By applying Equations~(\ref{eq:app:sec1:5}), (\ref{eq:app:sec2:1})-(\ref{eq:app:sec2:4}), and (\ref{eq:app:sec3:2}) to Equation~(\ref{eq:app:sec3:1}), we arrive at
\begin{align}
&\frac{1}{4\pi}\left[\left(\boldsymbol{\nabla}\times\left(\boldsymbol{\nabla}\times\left(\boldsymbol{\xi}\times\mathbf{B}_{\text{tor}}\right)\right)\right)\times\hat{\mathbf{B}}_{\text{tor}}+\left(\boldsymbol{\nabla}\times\mathbf{B}_{\text{tor}}\right)\times\left(\boldsymbol{\nabla}\times\left(\boldsymbol{\xi}\times\hat{\mathbf{B}}_{\text{tor}}\right)\right)\right.\notag\\
&\left.+\left(\boldsymbol{\xi}\cdot\boldsymbol{\nabla}\right)\left(\left(\boldsymbol{\nabla}\times\mathbf{B}_{\text{tor}}\right)\times\hat{\mathbf{B}}_{\text{tor}}\right)+\left(\boldsymbol{\nabla}\cdot\boldsymbol{\xi}\right)\left(\left(\boldsymbol{\nabla}\times\mathbf{B}_{\text{tor}}\right)\times\hat{\mathbf{B}}_{\text{tor}}\right)\right]\notag\\
=&\frac{1}{4\pi}\left[\left\{\left[\frac{2\Bphih\xir}{r}\delr{\Bphi}+2\Bphih\delr{\xir}\delr{\Bphi}+\frac{\Bphih}{r}\delth{\xith}\delr{\Bphi}+\frac{2\Bphih\Bphi}{r}\delr{\xir}+\Bphih\Bphi\delrsq{\xir}+\frac{\Bphih\Bphi}{r}\delrth{\xith}\right.\right.\right.\notag\\
&\left.+\frac{\Bphih\Bphi}{r^2\sin^2\theta}\delphisq{\xir}+\Bphih\xir\delrsq{\Bphi}+\frac{\Bphih}{r}\delr{\xith}\delth{\Bphi}+\frac{\Bphih\xith}{r}\delrth{\Bphi}\right]\mathbf{e}_{r}\notag\\
&+\left[\frac{\Bphih\xir}{r^2}\delth{\Bphi}+\frac{\Bphih}{r}\delr{\xir}\delth{\Bphi}+\frac{\Bphih\xith\cotth}{r^2}\delth{\Bphi}+\frac{2\Bphih}{r^2}\delth{\xith}\delth{\Bphi}+\frac{\Bphih\Bphi\xir\cotth}{r^2}\right.\notag\\
&+\frac{\Bphih\Bphi\cotth}{r}\delr{\xir}+\frac{\Bphih\Bphi}{r^2}\delth{\xir}+\frac{\Bphih\Bphi\cotth}{r^2}\delth{\xith}+\frac{\Bphih\Bphi}{r}\delrth{\xir}+\frac{\Bphih\Bphi}{r^2}\delthsq{\xith}\notag\\
&\left.\left.+\frac{\Bphih\Bphi}{r^2\sinthsq}\delphisq{\xith}+\frac{\Bphih}{r}\delth{\xir}\delr{\Bphi}+\frac{\Bphih\xir}{r}\delrth{\Bphi}+\frac{\Bphih\xir\cotth}{r}\delr{\Bphi}+\frac{\Bphih\xith}{r^2}\delthsq{\Bphi}\right]\mathbf{e}_{\theta}\right\}\notag\\
&+\left\{\left[\Bphih\delr{\Bphi}\delr{\xir}+\frac{\Bphih\xir}{r}\delr{\Bphi}+\frac{\Bphih}{r}\delr{\Bphi}\delth{\xith}+\xir\delr{\Bphi}\delr{\Bphih}+\frac{\xith}{r}\delr{\Bphi}\delth{\Bphih}\right.\right.\notag\\
&+\left.\frac{\Bphi\Bphih}{r}\delr{\xir}+\frac{\Bphi\Bphih\xir}{r^2}+\frac{\Bphi\Bphih}{r^2}\delth{\xith}+\frac{\xir\Bphi}{r}\delr{\Bphih}+\frac{\xith\Bphi}{r^2}\delth{\Bphih}\right]\mathbf{e}_r\notag\\
&+\left[\frac{\Bphi\Bphih\cotth}{r}\delr{\xir}+\frac{\Bphi\Bphih\cotth\xir}{r^2}+\frac{\Bphi\Bphih\cotth}{r^2}\delth{\xith}+\frac{\Bphi\cotth\xir}{r}\delr{\Bphih}+\frac{\Bphi\xith\cotth}{r^2}\delth{\Bphih}\right.\notag\\
&+\left.\frac{\Bphih}{r}\delth{\Bphi}\delr{\xir}+\frac{\Bphih\xir}{r^2}\delth{\Bphi}+\frac{\Bphih}{r^2} \delth{\Bphi}\delth{\xith}+\frac{\xir
}{r}\delth{\Bphi}\delr{\Bphih}+\frac{\xith}{r^2}\delth{\Bphi}\delth{\Bphih} \right]\mathbf{e}_{\theta}\notag\\
&\left.+\left[\frac{\Bphih}{r^2\sinth}\delth{\Bphi}\delphi{\xith}+\frac{\Bphi\Bphih\cotth}{r^2\sinth}\delphi{\xith}+\frac{\Bphih}{r\sinth}\delr{\Bphi} \delphi{\xir}+\frac{\Bphi\Bphih}{r^2\sinth}\delphi{\xir}\right]\mathbf{e}_{\phi}\right\}\notag\\
&+\left\{\left[\frac{\xir\Bphi\Bphih}{r^2}-\frac{\xir\Bphi}{r}\delr{\Bphih}-\frac{\xir\Bphih}{r}\delr{\Bphi} -\xir\delr{\Bphih}\delr{\Bphi}-\xir\Bphih\delrsq{\Bphi}-\frac{\xith\Bphi}{r^2}\delth{\Bphih} \right.\right.\notag\\
&-\left.\frac{\xith}{r}\delth{\Bphih}\delr{\Bphi}-\frac{\xith\Bphih}{r}\delrth{\Bphi}+\frac{\xith\Bphi\Bphih\cotth}{r^2}\right]\mathbf{e}_r\notag\\
&+\left[\frac{\xir\Bphi\Bphih\cotth}{r^2}-\frac{\xir\Bphi\cotth}{r}\delr{\Bphih}-\frac{\xir\Bphih\cotth}{r}\delr{\Bphi}+\frac{\xir\Bphih}{r^2}\delth{\Bphi}-\frac{\xir}{r}\delr{\Bphih}\delth{\Bphi}\right.\notag\\
&-\frac{\xir\Bphih}{r^2}\delrth{\Bphi}-\frac{\xith\Bphi\cotth}{r^2}\delth{\Bphih} -\frac{\xith\Bphih\cotth}{r^2}\delth{\Bphi}+\frac{\xith\Bphih\Bphi\cotthsq}{r^2}\notag\\
&\left.-\frac{\xith}{r^2}\delth{\Bphih}\delth{\Bphi}-\frac{\xith\Bphih}{r^2}\delthsq{\Bphi}-\frac{\xith\Bphih}{r}\delr{\Bphi}\right]\mathbf{e}_{\theta}\notag\\
&\left.+\left[-\frac{\xiphi\Bphi\Bphih}{r^2}-\frac{\xiphi\Bphih}{r}\delr{\Bphi}-\frac{\xiphi\Bphi\Bphih\cotthsq}{r^2}-\frac{\xiphi\Bphih\cotth}{r^2}\delth{\Bphi}\right]\mathbf{e}_{\phi}\right\}\notag\\
&+\left\{\left[-\frac{2\Bphi\Bphih\xir}{r^2}-\frac{\Bphi\Bphih}{r}\delr{\xir}-\frac{\Bphi\Bphih}{r^2}\delth{\xith}-\frac{\Bphi\Bphih\xith\cotth}{r^2}-\frac{\Bphi\Bphih}{r^2\sinth}\delphi{\xiphi}\right.\right.\notag\\
&\left.-\frac{2\Bphih\xir}{r}\delr{\Bphi}-\Bphih\delr{\xir}\delr{\Bphi}-\frac{\Bphih}{r}\delth{\xith}\delr{\Bphi}-\frac{\Bphih\xith\cotth}{r}\delr{\Bphi}-\frac{\Bphih}{r\sinth}\delphi{\xiphi}\delr{\Bphi}\right]\mathbf{e}_{r}\notag\\
&+\left[-\frac{2\Bphi\Bphih\xir\cotth}{r^2}-\frac{\Bphi\Bphih\cotth}{r}\delr{\xir}-\frac{\Bphi\Bphih\cotth}{r^2}\delth{\xith}-\frac{\Bphi\Bphih\xith\cotthsq}{r^2}-\frac{\Bphi\Bphih\cotth}{r^2\sinth}\delphi{\xiphi}\right.\notag\\
&\left.\left.\left.-\frac{2\Bphih\xir}{r^2}\delth{\Bphi}-\frac{\Bphih}{r}\delr{\xir}\delth{\Bphi}-\frac{\Bphih}{r^2}\delth{\xith}\delth{\Bphi}-\frac{\Bphih\xith\cotth}{r^2}\delth{\Bphi}-\frac{\Bphih}{r^2\sinth}\delphi{\xiphi}\delth{\Bphi}\right]\right\}\right]\mathbf{e}_{\theta}.\label{eq:app:sec3:3}
\end{align}
In this step, we kept the four terms from Equation~(\ref{eq:app:sec3:1}) separated and marked them off with curly brackets. We also applied the notation for the magnetic field from Section~\ref{sec:calc}.

The magnetic field components $\Bphi$ and $\Bphih$ are replaced by their representation 
\begin{align}
\Bphi &= a(r)\delth{}\gsharm{s}{0}{0}\left(\theta,\phi\right),\label{eq:app:sec3:21}\\
\Bphih &= \hat{a}(r)\delth{}\gsharm{s\p}{0}{0}\left(\theta,\phi\right),\label{eq:app:sec3:22}
\end{align}
and the components of the eigenfunctions are expressed as given by Equation~(\ref{eq:sec1:14}).

In the next step, the eigenfunction $\overline{\hat{\boldsymbol{\xi}}}_{l\p,m\p}\left(r,\theta,\phi\right)$, where the bar denotes complex conjugation, is multiplied from the left, and the integral over the whole stellar volume is applied. As the radial and horizontal displacement amplitudes are purely real, the complex conjugation on them can be dropped: $\overline{\hat{\xi}^r(r)}=\hat{\xi}^r(r)$, $\overline{\hat{\xi}^h(r)}=\hat{\xi}^h(r)$. Then, terms that carry only a radial dependence are separated from those with an angular dependence. Terms with identical angular dependence can be factored out and summarized. After all of these steps, we finally arrive at Equation~(\ref{eq:sec4:3}). The factors with an angular dependence, $\mathcal{T}_i$, with $i=1,\dots,25$, which appear in that equation, are listed in Appendix~\ref{sec:app:kernels}.

\section{The Generalized Spherical Harmonics and Their Properties}\label{sec:app:gsh}
The generalized spherical harmonics can be used to expand tensor fields of any order. Here, they are employed in the representation of the eigenfunctions and the toroidal magnetic fields as given in Equations~(\ref{eq:sec1:14}), (\ref{eq:sec4:1}), and (\ref{eq:sec4:2}). We follow the conventions for the generalized spherical harmonics of \cite{Dahlen1998}, who largely build upon the notation and sign conventions from \cite{Phinney1973}. 

The scalar generalized spherical harmonics are defined by 
\begin{align}
\gsharm{l}{N}{m} = \gamma_{l} P_{lm}^N\left(\costh\right) e^{im\phi}\label{app:eq1},
\end{align}
where $\gamma_{l} = \sqrt{\left(2l+1\right)/4\pi}$ and $P_{lm}^N\left(\costh\right)$ are the generalized Legendre functions. In this convention, $\sharm{l}{m}=\gsharm{l}{0}{m}$ holds, where $\sharm{l}{m}$ is an ordinary spherical harmonic function of degree $l$ and azimuthal order $m$. 

We collect all identities and relations of the generalized spherical harmonics, which are needed in the calculation of the matrix element $H_{k\p k}$, especially in the calculation of volume integrals over products of these functions; see Appendix~\ref{sec:app:integral} and \ref{sec:app:kernels}. 

The generalized spherical harmonics satisfy the recursion relation
\begin{align}
\frac{N\cos\theta -m}{\sin\theta}\gsharm{l}{N}{m} =& \frac{1}{\sqrt{2}} \left(\om{N+1}{l}\gsharm{l}{N+1}{m} +\om{N}{l}\gsharm{l}{N-1}{m}\right),\label{app:eq2}
\end{align}
with
\begin{align}
\om{N}{l} = \sqrt{\frac{\left(l+N\right)\left(l-N+1\right)}{2}}\label{app:eq3.1}.
\end{align}
The colatitudinal derivative of $\gsharm{l}{N}{m}$ can be expressed in terms of $\gsharm{l}{N\pm1}{m}$:
\begin{align}
\delth{}\gsharm{l}{N}{m} &= \frac{1}{\sqrt{2}}\left(\om{N}{l}\gsharm{l}{N-1}{m}-\om{N+1}{l}\gsharm{l}{N+1}{m}\right)\label{app:eq4}.        
\end{align}
From Equation~(\ref{app:eq1}), it can be seen that the azimuthal derivative is given by
\begin{align}
\delphi{}\gsharm{l}{N}{m} &= im\gsharm{l}{N}{m}\label{app:eq6}.
\end{align}
Higher colatitudinal and azimuthal derivatives can easily be constructed by an iterative application of Equation~(\ref{app:eq4})-(\ref{app:eq6}). For example, the second colatitudinal derivative of $\gsharm{l}{N}{m}$ is given by
\begin{align}
\delthsq{}\gsharm{l}{N}{m} &=\frac{1}{2}\left(\om{N}{l}\om{N-1}{l}\gsharm{l}{N-2}{m}-\om{N}{l}\om{N}{l}\gsharm{l}{N}{m}-\om{N+1}{l}\om{N+1}{l}\gsharm{l}{N}{m}+\om{N+1}{l}\om{N+2}{l}\gsharm{l}{N+2}{m}\right)\label{app:eq7}.
\end{align}
The complex conjugate of the $\gsharm{l}{N}{m}$ is 
\begin{align}
\overline{\gsharm{l}{N}{m}} = \left(-1\right)^{-N-m}\gsharm{l}{-N}{-m}\label{app:eq8}.
\end{align}
\subsection{Elimination of Trigonometric Functions}
Derivatives and trigonometric functions appear in the calculation of the angular integral of the angle-dependent factors of the matrix element; see Appendix~\ref{sec:app:kernels}. The derivatives can be expressed via relations (\ref{app:eq4}) and (\ref{app:eq8}), and their iterative application. The trigonometric functions can be rewritten in terms of generalized spherical harmonics with the help of Equation~(\ref{app:eq2}). Setting one of the upper indices equal to zero yields a relation for eliminating a factor $\sin^{-1}\theta$ (if $N=0$) or for eliminating a factor $\cotth$ (if $m=0$). 

Two more useful relations can be obtained by taking the colatitudinal derivative of the recursion relation (\ref{app:eq2}), which yields
\begin{align}
&\left(-\frac{N}{\sinthsq}+\frac{m\cotth}{\sinth}\right)\gsharm{l}{N}{m}+ \left(\frac{N\cotth}{\sqrt{2}}-\frac{m}{\sqrt{2}\sinth}\right)\left(\om{N}{l}\gsharm{l}{N-1}{m}-\om{N+1}{l}\gsharm{l}{N+1}{m}\right) \notag\\
&= \frac{1}{2} \left(\om{N+1}{l}\om{N+1}{l}\gsharm{l}{N}{m}-\om{N+1}{l}\om{N+2}{l}\gsharm{l}{N+2}{m} +\om{N}{l}\om{N-1}{l}\gsharm{l}{N-2}{m}-\om{N}{l}\om{N}{l}\gsharm{l}{N}{m}\right).\label{app:eq8a}
\end{align}
By setting $N=0$ in Equation~(\ref{app:eq8a}) and using $\gsharm{1}{1}{0} = \gamma_1 P_1^{1,0} =  \sqrt{\frac{3}{4\pi}}\frac{1}{\sqrt{2}}\sinth$, we obtain
\begin{align}
m\cotth\gsharm{l}{0}{m}= \sqrt{\frac{2\pi}{3}}\om{0}{l}\om{2}{l} &\left(-\gsharm{l}{2}{m}\gsharm{1}{1}{0} +\gsharm{l}{-2}{m}\gsharm{1}{1}{0}\right)+\frac{m}{\sqrt{2}}\om{0}{l}\left(\gsharm{l}{-1}{m}-\gsharm{l}{1}{m}\right)\label{app:eq8b}.
\end{align}
By setting $m=0$ in Equation~(\ref{app:eq8a}) and using $\gsharm{1}{0}{0} = \gamma_1 P_1^{0,0} =  \sqrt{\frac{3}{4\pi}}\costh$, we obtain
\begin{align}
\frac{N}{\sinth}\gsharm{l}{N}{0} = -\sqrt{\frac{2\pi}{3}} &\left(\om{N+1}{l}\om{N+1}{l}\gsharm{l}{N}{0}\gsharm{1}{1}{0}-\om{N+1}{l}\om{N+2}{l}\gsharm{l}{N+2}{0}\gsharm{1}{1}{0}\right.\notag\\ &\left.+\om{N}{l}\om{N-1}{l}\gsharm{l}{N-2}{0}\gsharm{1}{1}{0}-\om{N}{l}\om{N}{l}\gsharm{l}{N}{0}\gsharm{1}{1}{0}\right)\notag\\
&+N\sqrt{\frac{2\pi}{3}}\left(\om{N}{l}\gsharm{l}{N-1}{0}\gsharm{1}{0}{0}-\om{N+1}{l}\gsharm{l}{N+1}{0}\gsharm{1}{0}{0}\right)\label{app:eq8c}.
\end{align}

\section{The Wigner $3j$ Symbol}\label{sec:app:wtj}
The Wigner $3j$ symbols are maximally symmetrized representations of the Clebsch-Gordon coefficients. They represent the coupling of three angular momenta to a resultant zero momentum \citep{Racah1942, Edmonds}. The most symmetrical form to calculate them is
\begin{align}
\wtj{l_1}{l_2}{l_3}{m_1}{m_2}{m_3} =& \left(-1\right)^{l_2+l_3+m_1}\frac{\sqrt{\left(l_1+l_2-l_3\right)!\left(l_1-l_2+l_3\right)!\left(-l_1+l_2+l_3\right)!}}{\sqrt{\left(l_1+l_2+l_3+1\right)!}}\notag\\
&\cdot\sqrt{\left(l_1+m_1\right)!\left(l_1-m_1\right)!\left(l_2+m_2\right)!\left(l_2-m_2\right)!\left(l_3+m_3\right)!\left(l_3-m_3\right)!}\sum_{k}\frac{\left(-1\right)^k}{D_k},\label{app:eq:wtj1}
\end{align}
where $k\in\mathbb{N}_0$ runs over all values for which $D_k\ne0$ with
\begin{align}
D_k = k!\left(-l_1+l_2+l_3-k\right)!\left(l_2-m_2-k\right)!\left(l_3+m_3-k\right)!\left(l_1-l_2-m_3+k\right)!\left(l_1-l_3+m_2+k\right)!.
\end{align}
In general, $l_i$ and $m_i$ can be integers or half-integers. In this work, $l_i$ are always non-negative integers and $m_i$ are integers in the range $-l_i \le m_i \le l_i$. 

We list only those properties of the Wigner $3j$ symbols that are useful in the calculation of the general matrix element $H_{k\p k}$. More extensive lists can be found in, e.g., \cite{Dahlen1998}, \cite{Edmonds}, and \cite{Regge1958}.

The numerical value of a $3j$ symbol is unchanged under an even permutation of its columns:
\begin{align}
\wtj{l_1}{l_2}{l_3}{m_1}{m_2}{m_3} = \wtj{l_2}{l_3}{l_1}{m_2}{m_3}{m_1} = \wtj{l_3}{l_1}{l_2}{m_3}{m_1}{m_2}\label{app:eq:wtj2}.
\end{align}
An odd permutation of its columns is equivalent to multiplication by $\left(-1\right)^{l_1+l_2+l_3}$:
\begin{align}
\wtj{l_2}{l_1}{l_3}{m_2}{m_1}{m_3} = \wtj{l_1}{l_3}{l_2}{m_1}{m_3}{m_2} = \wtj{l_3}{l_2}{l_1}{m_3}{m_2}{m_1} = \left(-1\right)^{l_1+l_2+l_3}\wtj{l_1}{l_2}{l_3}{m_1}{m_2}{m_3}\label{app:eq:wtj3}.
\end{align}
Selection rules for the allowed ranges of the $l_i$ can be constructed from the following properties:
\begin{align}
\wtj{l_1}{l_2}{l_3}{m_1}{m_2}{m_3} =& 0, \text{ if $m_1=m_2=m_3=0$ and $l_1+l_2+l_3$ is odd},\label{app:eq:wtj4}
\end{align}
\begin{subnumcases}{\wtj{l_1}{l_2}{l_3}{m_1}{m_2}{m_3}\ne 0,\text{ only if }}
&$m_1+m_2+m_3=0$,\label{app:eq:wtj5.1}\\
&\text{and }$\left|m_i\right|\le l_i$\text{ for }$i=1,2,3,$\label{app:eq:wtj5.2}\\
&\text{and }$\left|l_i-l_j\right|\le l_k$\text{ for }$i,j,k=1,2,3$\label{app:eq:wtj5.3}.
\end{subnumcases}

\section{Integral of the Products of Generalized Spherical Harmonics on the Unit Sphere}\label{sec:app:integral}
To calculate the analytical expression of the general matrix element, it is necessary to integrate the product of four, five, and six generalized spherical harmonics. In \cite{Dahlen1998}, it is shown that the integral of a product of three generalized spherical harmonics on the unit sphere can be written as
\begin{align}
\int\overline{\gsharm{l}{N}{m}}\gsharm{l_1}{N_1}{m_1}\gsharm{l_2}{N_2}{m_2}\D \Omega =4\pi\left(-1\right)^{N+m}\gamma_{l}\gamma_{l_1}\gamma_{l_2}\wtj{l}{l_1}{l_2}{-N}{N_1}{N_2}\wtj{l}{l_1}{l_2}{-m}{m_1}{m_2}\label{app:eq9},
\end{align}
where $\gamma_{k} = \sqrt{\left(2k+1\right)/4\pi}$.\\
The product of two generalized spherical harmonics is given by
\begin{align}
\gsharm{l_1}{N_1}{m_1}\gsharm{l_2}{N_2}{m_2}= \sum_{l,N,m}4\pi\left(-1\right)^{N+m}\gamma_{l}\gamma_{l_1}\gamma_{l_2}\wtj{l_1}{l_2}{l}{N_1}{N_2}{-N}\wtj{l_1}{l_2}{l}{m_1}{m_2}{-m}\gsharm{l}{N}{m}\label{app:eq10}.
\end{align}
We can use Equations~(\ref{app:eq9})-(\ref{app:eq10}) to calculate the integral over four generalized spherical harmonics:
\begin{align}
&\int\overline{\gsharm{l_1}{N_1}{m_1}}\gsharm{l_2}{N_2}{m_2}\gsharm{l_3}{N_3}{m_3}\gsharm{l_4}{N_4}{m_4}\D \Omega\notag\\
&=4\pi\gamma_{l_1}\gamma_{l_2}\gamma_{l_3}\gamma_{l_4}\sum\limits_{l,N,m}\left(-1\right)^{N+N_1+m+m_1}\left(2l+1\right)\wtj{l_1}{l_2}{l}{-N_1}{N_2}{N}\wtj{l_1}{l_2}{l}{-m_1}{m_2}{m}\notag\\
&\cdot\wtj{l_3}{l_4}{l}{N_3}{N_4}{-N}\wtj{l_3}{l_4}{l}{m_3}{m_4}{-m}
\label{app:eq11}.
\end{align}
The sum extends over all values of $l,N$, and $m$, which are allowed according to Equations~(\ref{app:eq:wtj5.1})-(\ref{app:eq:wtj5.3}).

By employing Equation~(\ref{app:eq:wtj5.1}), we can deduce two useful conditions for the indices of the four generalized spherical harmonics in Equation~(\ref{app:eq11}) if the integral shall not vanish:
\begin{align}
N_1 &= \sum_{i>1} N_i\label{app:eq:sumN},\\
m_1 &= \sum_{i>1} m_i\label{app:eq:summ}.
\end{align}

By iterative application of Equations~(\ref{app:eq9})-(\ref{app:eq11}), also integrals over products of five, six, or any number of generalized spherical harmonics can also be calculated. Conditions (\ref{app:eq:sumN}) and (\ref{app:eq:summ}) also hold for integrals over products with more than four generalized spherical harmonics. 

\section{The angular kernels for toroidal magnetic fields}\label{sec:app:kernels}
In the general matrix element, given by Equation~(\ref{eq:sec4:3}), we factored out the parts that hold an angular dependence. These factors are given by the generalized spherical harmonic functions from the two eigenfunctions and the two magnetic field components, which appear in the Lorentz force. They also contain trigonometric functions, which appear due to the derivatives in spherical geometry, and derivatives acting on the generalized spherical harmonics. In this list, we include the factors $\mathcal{T}_{13}$-$\mathcal{T}_{16}$ and $\mathcal{T}_{21}$. In our derivations, they appear for individual terms during the steps outlined in Appendix~\ref{sec:app:calc} but cancel along the way to the final result. We chose to list them here, because they may be of interest for studies that focus on the Eulerian perturbation of the Lorentz force or if poloidal magnetic fields are included. The individual factors are given by
\begin{align}
\mathcal{T}_1 &= \overline{\gsharm{l\p}{0}{m\p}} \gsharm{l}{0}{m}\delth{}\gsharm{s}{0}{0}\delth{}\gsharm{s\p}{0}{0}\label{app:eq:kernel:deriv:1}\\
\mathcal{T}_2 &= \overline{\gsharm{l\p}{0}{m\p}} \delthsq{}\gsharm{l}{0}{m}\delth{}\gsharm{s}{0}{0}\delth{}\gsharm{s\p}{0}{0}\label{app:eq:kernel:deriv:2}\\
\mathcal{T}_3 &= \frac{1}{\sin^2\theta}\overline{\gsharm{l\p}{0}{m\p}}\delphisq{}\gsharm{l}{0}{m}\delth{}\gsharm{s}{0}{0}\delth{}\gsharm{s\p}{0}{0}\label{app:eq:kernel:deriv:3}\\
\mathcal{T}_4 &= \overline{\gsharm{l\p}{0}{m\p}}\delth{}\gsharm{l}{0}{m}\delthsq{}\gsharm{s}{0}{0}\delth{}\gsharm{s\p}{0}{0}\label{app:eq:kernel:deriv:4}\\
\mathcal{T}_5 &= \delth{}\overline{\gsharm{l\p}{0}{m\p}}\gsharm{l}{0}{m}\delthsq{}\gsharm{s}{0}{0}\delth{}\gsharm{s\p}{0}{0}\label{app:eq:kernel:deriv:5}\\
\mathcal{T}_6 &= \cotth\delth{}\overline{\gsharm{l\p}{0}{m\p}}\delth{}\gsharm{l}{0}{m} \delthsq{}\gsharm{s}{0}{0}\delth{}\gsharm{s\p}{0}{0}\label{app:eq:kernel:deriv:6}\\
\mathcal{T}_7 &= \delth{}\overline{\gsharm{l\p}{0}{m\p}}\delthsq{}\gsharm{l}{0}{m}\delthsq{}\gsharm{s}{0}{0}\delth{}\gsharm{s\p}{0}{0}\label{app:eq:kernel:deriv:7}\\
\mathcal{T}_8 &= \cotth\delth{}\overline{\gsharm{l\p}{0}{m\p}}\gsharm{l}{0}{m}\delth{}\gsharm{s}{0}{0}\delth{}\gsharm{s\p}{0}{0}\label{app:eq:kernel:deriv:8}\\
\mathcal{T}_9 &= \cotth\delth{}\overline{\gsharm{l\p}{0}{m\p}}\delthsq{}\gsharm{l}{0}{m}\delth{}\gsharm{s}{0}{0} \delth{}\gsharm{s\p}{0}{0}\label{app:eq:kernel:deriv:9}\\
\mathcal{T}_{10} &= \delth{}\overline{\gsharm{l\p}{0}{m\p}}\delth{}\gsharm{l}{0}{m}\delth{}\gsharm{s}{0}{0}\delth{}\gsharm{s\p}{0}{0}\label{app:eq:kernel:deriv:10}\\
\mathcal{T}_{11} &= \delth{}\overline{\gsharm{l\p}{0}{m\p}} \frac{\partial^3}{\partial\theta^3}\gsharm{l}{0}{m} \delth{}\gsharm{s}{0}{0}\delth{}\gsharm{s\p}{0}{0}\label{app:eq:kernel:deriv:11}\\
\mathcal{T}_{12} &= \frac{1}{\sin^2\theta}\delth{}\overline{\gsharm{l\p}{0}{m\p}}\delphisq{}\delth{}\gsharm{l}{0}{m}\delth{}\gsharm{s}{0}{0}\delth{}\gsharm{s\p}{0}{0}\label{app:eq:kernel:deriv:12}\\
\mathcal{T}_{13} &= \delth{}\overline{\gsharm{l\p}{0}{m\p}}\delth{}\gsharm{l}{0}{m}\frac{\partial^3}{\partial\theta^3}\gsharm{s}{0}{0}\delth{}\gsharm{s\p}{0}{0}\label{app:eq:kernel:deriv:13}\\
\mathcal{T}_{14} &= \overline{\gsharm{l\p}{0}{m\p}}\delth{}\gsharm{l}{0}{m}\delth{}\gsharm{s}{0}{0} \delthsq{}\gsharm{s\p}{0}{0}\label{app:eq:kernel:deriv:14}\\
\mathcal{T}_{15} &= \cotth\delth{}\overline{\gsharm{l\p}{0}{m\p}}\delth{}\gsharm{l}{0}{m}\delth{}\gsharm{s}{0}{0} \delthsq{}\gsharm{s\p}{0}{0}\label{app:eq:kernel:deriv:15}\\
\mathcal{T}_{16} &= \delth{}\overline{\gsharm{l\p}{0}{m\p}}\delth{}\gsharm{l}{0}{m} \delthsq{}\gsharm{s}{0}{0} \delthsq{}\gsharm{s}{0}{0}\label{app:eq:kernel:deriv:16}\\
\mathcal{T}_{17} &= \frac{1}{\sinthsq}\delphi{}\overline{\gsharm{l\p}{0}{m\p}} \delthphi{}\gsharm{l}{0}{m}\delthsq{}\gsharm{s}{0}{0} \delth{}\gsharm{s\p}{0}{0}\label{app:eq:kernel:deriv:17}\\
\mathcal{T}_{18} &= \frac{\cotth}{\sinthsq}\delphi{}\overline{\gsharm{l\p}{0}{m\p}} \delthphi{}\gsharm{l}{0}{m}\delth{}\gsharm{s}{0}{0}\delth{}\gsharm{s\p}{0}{0}\label{app:eq:kernel:deriv:18}\\
\mathcal{T}_{19} &= \frac{1}{\sinthsq}\delphi{}\overline{\gsharm{l\p}{0}{m\p}} \delphi{} \gsharm{l}{0}{m}\delth{}\gsharm{s}{0}{0}\delth{}\gsharm{s\p}{0}{0}\label{app:eq:kernel:deriv:19}\\
\mathcal{T}_{20} &= \cotth\overline{\gsharm{l\p}{0}{m\p}} \delth{}\gsharm{l}{0}{m}\delth{}\gsharm{s}{0}{0}\delth{}\gsharm{s\p}{0}{0}\label{app:eq:kernel:deriv:20}\\
\mathcal{T}_{21} &= \cotthsq\delth{}\overline{\gsharm{l\p}{0}{m\p}}\delth{}\gsharm{l}{0}{m} \delth{}\gsharm{s}{0}{0}\delth{}\gsharm{s\p}{0}{0} \label{app:eq:kernel:deriv:21}\\
\mathcal{T}_{22} &= \frac{\cotthsq}{\sinthsq}\delphi{}\overline{\gsharm{l\p}{0}{m\p}} \delphi{}\gsharm{l}{0}{m}\delth{}\gsharm{s}{0}{0}\delth{}\gsharm{s\p}{0}{0}\label{app:eq:kernel:deriv:22}\\
\mathcal{T}_{23} &= \frac{\cotth}{\sinthsq}\delphi{}\overline{\gsharm{l\p}{0}{m\p}}\delphi{}\gsharm{l}{0}{m}\delthsq{}\gsharm{s}{0}{0}\delth{}\gsharm{s\p}{0}{0}\label{app:eq:kernel:deriv:23}\\
\mathcal{T}_{24} &= \frac{\cotth}{\sinthsq}\delth{}\overline{\gsharm{l\p}{0}{m\p}}\delphisq{}\gsharm{l}{0}{m}\delth{}\gsharm{s}{0}{0} \delth{}\gsharm{s\p}{0}{0}\label{app:eq:kernel:deriv:24}\\
\mathcal{T}_{25} &= \frac{1}{\sinthsq}\delth{}\overline{\gsharm{l\p}{0}{m\p}}\delphisq{}\gsharm{l}{0}{m}\delthsq{}\gsharm{s}{0}{0}\delth{}\gsharm{s\p}{0}{0}\label{app:eq:kernel:deriv:25}
\end{align}

As the integral in Equation~(\ref{eq:sec4:3}) extends over the whole volume and we were able to separate the radial from the angular parts, we can carry out the radial and angular integrals individually. In the following, we give the explicit expressions for the angular kernels~(\ref{app:eq:kernel:deriv:1})-(\ref{app:eq:kernel:deriv:25}). In the derivation of Equations~(\ref{app:eq:kernel:fin:1})-(\ref{app:eq:kernel:fin:25}), Equations~(\ref{app:eq2})-(\ref{app:eq7}) and (\ref{app:eq8c}) are used to eliminate the derivatives of the generalized spherical harmonics and the trigonometric functions. We utilize selection rule (\ref{app:eq:sumN}) to eliminate terms whose angular integral is equal to zero for any combination of harmonic degrees and azimuthal orders of the eigenfunctions and the magnetic field components. To ease readability, we use that
\begin{align}
\om{-N}{l} &= \om{N+1}{l},
\end{align}
which can be seen from Equation~(\ref{app:eq3.1}). Also, we set $m'=m$, which is required by Equation~(\ref{app:eq:summ}). With all these, we find after considerable but straightforward algebra
\begin{align}	
\mathcal{T}_{1}=&=-\frac{1}{2}\left(\om{0}{s}\om{0}{s\p}\overline{\gsharm{l\p}{0}{m}}\gsharm{l}{0}{m}\gsharm{s}{-1}{0}\gsharm{s\p}{1}{0}
+\om{0}{s}\om{0}{s\p}\overline{\gsharm{l\p}{0}{m}}\gsharm{l}{0}{m}\gsharm{s}{1}{0}\gsharm{s\p}{-1}{0}\right)\label{app:eq:kernel:fin:1}\\
\mathcal{T}_{2}=&=\frac{1}{4}\om{0}{s\p}\om{0}{s}\om{0}{l}\left(
\om{2}{l}\overline{\gsharm{l\p}{0}{m}}\gsharm{l}{-2}{m}\gsharm{s}{1}{0}\gsharm{s\p}{1}{0}
+2\om{0}{l}\overline{\gsharm{l\p}{0}{m}}\gsharm{l}{0}{m}\gsharm{s}{-1}{0}\gsharm{s\p}{1}{0}\right.\notag\\
&+2\om{0}{l}\overline{\gsharm{l\p}{0}{m}}\gsharm{l}{0}{m}\gsharm{s}{1}{0}\gsharm{s\p}{-1}{0}
\left.+\om{2}{l}\overline{\gsharm{l\p}{0}{m}}\gsharm{l}{2}{m}\gsharm{s}{-1}{0}\gsharm{s\p}{-1}{0}\right)\label{app:eq:kernel:fin:2}\\
\mathcal{T}_{3}=&-\frac{1}{4}\om{0}{s}\om{0}{s\p}\om{0}{l\p}\om{0}{l}\left(\overline{\gsharm{l\p}{-1}{m}}\gsharm{l}{1}{m}\gsharm{s}{-1}{0}\gsharm{s\p}{-1}{0}
-\overline{\gsharm{l\p}{1}{m}}\gsharm{l}{1}{m}\gsharm{s}{-1}{0}\gsharm{s\p}{1}{0}\right.\notag\\ 
&-\overline{\gsharm{l\p}{-1}{m}}\gsharm{l}{-1}{m}\gsharm{s}{-1}{0}\gsharm{s\p}{1}{0}	-\overline{\gsharm{l\p}{1}{m}}\gsharm{l}{1}{m}\gsharm{s}{1}{0}\gsharm{s\p}{-1}{0}\notag\\ 
&\left.-\overline{\gsharm{l\p}{-1}{m}}\gsharm{l}{-1}{m}\gsharm{s}{1}{0}\gsharm{s\p}{-1}{0}
+\overline{\gsharm{l\p}{1}{m}}\gsharm{l}{-1}{m}\gsharm{s}{1}{0}\gsharm{s\p}{1}{0}\right)\label{app:eq:kernel:fin:3}\\
\mathcal{T}_{4}=&\frac{1}{4}\om{0}{s}\om{0}{s\p}\om{0}{l}\left(\om{2}{s}\overline{\gsharm{l\p}{0}{m}}\gsharm{l}{-1}{m}\gsharm{s}{2}{0}\gsharm{s\p}{-1}{0}\right.+2\om{0}{s}\overline{\gsharm{l\p}{0}{m}}\gsharm{l}{-1}{m}\gsharm{s}{0}{0}\gsharm{s\p}{1}{0}\notag\\
&+2\om{0}{s}\overline{\gsharm{l\p}{0}{m}}\gsharm{l}{1}{m}\gsharm{s}{0}{0}\gsharm{s\p}{-1}{0}
\left.+\om{2}{s}\overline{\gsharm{l\p}{0}{m}}\gsharm{l}{1}{m}\gsharm{s}{-2}{0}\gsharm{s\p}{1}{0}\right)\label{app:eq:kernel:fin:4}\\
\mathcal{T}_{5}=& -\frac{1}{4}\om{0}{s}\om{0}{s\p}\om{0}{l\p}\left(2\om{0}{s}\overline{\gsharm{l\p}{-1}{m}}\gsharm{l}{0}{m}\gsharm{s}{0}{0}\gsharm{s\p}{-1}{0} +\om{2}{s}\overline{\gsharm{l\p}{-1}{m}}\gsharm{l}{0}{m}\gsharm{s}{-2}{0}\gsharm{s\p}{1}{0}\right.\notag\\
&\left.+\om{2}{s}\overline{\gsharm{l\p}{1}{m}}\gsharm{l}{0}{m}\gsharm{s}{2}{0}\gsharm{s\p}{-1}{0}
+2\om{0}{s}\overline{\gsharm{l\p}{1}{m}}\gsharm{l}{0}{m}\gsharm{s}{0}{0}\gsharm{s\p}{1}{0}\right)\label{app:eq:kernel:fin:5}\\
\mathcal{T}_{6}=&\frac{1}{8}\om{0}{s}\om{0}{s\p}\om{0}{l\p}\om{0}{l}\left(
4\om{0}{s}\om{0}{s\p}\overline{\gsharm{l\p}{-1}{m}}\gsharm{l}{-1}{m}\gsharm{s}{0}{0} \gsharm{s\p}{0}{0}\right.\notag\\
&-\om{2}{s}\om{2}{s\p}\overline{\gsharm{l\p}{-1}{m}}\gsharm{l}{-1}{m}\gsharm{s}{2}{0} \gsharm{s\p}{-2}{0}
-\om{2}{s}\om{2}{s\p}\overline{\gsharm{l\p}{-1}{m}}\gsharm{l}{-1}{m}\gsharm{s}{-2}{0} \gsharm{s\p}{2}{0}\notag\\
&+2\om{2}{s}\om{0}{s\p}\overline{\gsharm{l\p}{1}{m}}\gsharm{l}{-1}{m}\gsharm{s}{2}{0} \gsharm{s\p}{0}{0}
-2\om{0}{s}\om{2}{s\p}\overline{\gsharm{l\p}{1}{m}}\gsharm{l}{-1}{m}\gsharm{s}{0}{0} \gsharm{s\p}{2}{0}\notag\\
&+2\om{2}{s}\om{0}{s\p}\overline{\gsharm{l\p}{-1}{m}}\gsharm{l}{1}{m}\gsharm{s}{-2}{0} \gsharm{s\p}{0}{0}
-2\om{0}{s}\om{2}{s\p}\overline{\gsharm{l\p}{-1}{m}}\gsharm{l}{1}{m}\gsharm{s}{0}{0} \gsharm{s\p}{-2}{0}\notag\\
&-\om{2}{s}\om{2}{s\p}\overline{\gsharm{l\p}{1}{m}}\gsharm{l}{1}{m}\gsharm{s}{2}{0} \gsharm{s\p}{-2}{0}
-\om{2}{s}\om{2}{s\p}\overline{\gsharm{l\p}{1}{m}}\gsharm{l}{1}{m}\gsharm{s}{-2}{0} \gsharm{s\p}{2}{0}\notag\\
&+\left.4\om{0}{s}\om{0}{s\p}\overline{\gsharm{l\p}{1}{m}}\gsharm{l}{1}{m}\gsharm{s}{0}{0} \gsharm{s\p}{0}{0}\right)\label{app:eq:kernel:fin:6}\\
\mathcal{T}_{7}=&\frac{1}{8}\om{0}{s}\om{0}{s\p}\om{0}{l\p}\om{0}{l}\left(\om{2}{s}\om{2}{l}\overline{\gsharm{l\p}{-1}{m}}\gsharm{l}{-2}{m}\gsharm{s}{2}{0}\gsharm{s\p}{-1}{0}
+2\om{0}{s}\om{2}{l}\overline{\gsharm{l\p}{-1}{m}}\gsharm{l}{-2}{m}\gsharm{s}{0}{0}\gsharm{s\p}{1}{0}\right.\notag\\
&+4\om{0}{s}\om{0}{l}\overline{\gsharm{l\p}{-1}{m}}\gsharm{l}{0}{m}\gsharm{s}{0}{0}\gsharm{s\p}{-1}{0}
+2\om{2}{s}\om{0}{l}\overline{\gsharm{l\p}{-1}{m}}\gsharm{l}{0}{m}\gsharm{s}{-2}{0}\gsharm{s\p}{1}{0}\notag\\
& +\om{2}{s}\om{2}{l}\overline{\gsharm{l\p}{-1}{m}}\gsharm{l}{2}{m}\gsharm{s}{-2}{0}\gsharm{s\p}{-1}{0}
+\om{2}{s}\om{2}{l}\overline{\gsharm{l\p}{1}{m}}\gsharm{l}{-2}{m}\gsharm{s}{2}{0}\gsharm{s\p}{1}{0}\notag\\
&+2\om{2}{s}\om{0}{l}\overline{\gsharm{l\p}{1}{m}}\gsharm{l}{0}{m}\gsharm{s}{2}{0}\gsharm{s\p}{-1}{0}
+4\om{0}{s}\om{0}{l}\overline{\gsharm{l\p}{1}{m}}\gsharm{l}{0}{m}\gsharm{s}{0}{0}\gsharm{s\p}{1}{0}\notag\\
&\left.+2\om{0}{s}\om{2}{l}\overline{\gsharm{l\p}{1}{m}}\gsharm{l}{2}{m}\gsharm{s}{0}{0}\gsharm{s\p}{-1}{0}
+\om{2}{s}\om{2}{l}\overline{\gsharm{l\p}{1}{m}}\gsharm{l}{2}{m}\gsharm{s}{-2}{0}\gsharm{s\p}{1}{0}\right)\label{app:eq:kernel:fin:7}\\
\mathcal{T}_{8}=& \frac{1}{4}\om{0}{s}\om{0}{s\p}\om{0}{l\p}\left(
-2\om{0}{s\p}\overline{\gsharm{l\p}{-1}{m}}\gsharm{l}{0}{m}\gsharm{s}{-1}{0}\gsharm{s\p}{0}{0}
+\om{2}{s\p}\overline{\gsharm{l\p}{-1}{m}}\gsharm{l}{0}{m}\gsharm{s}{1}{0}\gsharm{s\p}{-2}{0}\right.\notag\\
&\left. +\om{2}{s\p} \overline{\gsharm{l\p}{1}{m}}\gsharm{l}{0}{m}\gsharm{s}{-1}{0}\gsharm{s\p}{2}{0}
-2\om{0}{s\p} \overline{\gsharm{l\p}{1}{m}}\gsharm{l}{0}{m}\gsharm{s}{1}{0}\gsharm{s\p}{0}{0}\right)\label{app:eq:kernel:fin:8}\\
\mathcal{T}_{9}=&\frac{1}{8}\om{0}{s}\om{0}{s\p}\om{0}{l\p}\om{0}{l}\left(
2\om{0}{s\p}\om{0}{l}\overline{\gsharm{l\p}{-1}{m}}\gsharm{l}{0}{m}\gsharm{s}{-1}{0}\gsharm{s\p}{0}{0}
-\om{2}{s\p}\om{2}{l}\overline{\gsharm{l\p}{-1}{m}}\gsharm{l}{2}{m}\gsharm{s}{-1}{0}\gsharm{s\p}{-2}{0}\right.\notag\\
&+\om{0}{s\p}\om{2}{l}\overline{\gsharm{l\p}{1}{m}}\gsharm{l}{2}{m}\gsharm{s}{-1}{0}\gsharm{s\p}{0}{0}
+\om{0}{s\p}\om{2}{l}\overline{\gsharm{l\p}{-1}{m}}\gsharm{l}{-2}{m}\gsharm{s}{1}{0}\gsharm{s\p}{0}{0}\notag\\
&-2\om{2}{s\p}\om{0}{l}\overline{\gsharm{l\p}{-1}{m}}\gsharm{l}{0}{m}\gsharm{s}{1}{0}\gsharm{s\p}{-2}{0}
+2\om{0}{s\p}\om{0}{l}\overline{\gsharm{l\p}{1}{m}}\gsharm{l}{0}{m}\gsharm{s}{1}{0}\gsharm{s\p}{0}{0}\notag\\
&-\om{2}{s\p}\om{2}{l}\overline{\gsharm{l\p}{1}{m}}\gsharm{l}{2}{m}\gsharm{s}{1}{0}\gsharm{s\p}{-2}{0}
-\om{0}{s\p}\om{2}{l}\overline{\gsharm{l\p}{-1}{m}}\gsharm{l}{-2}{m}\gsharm{s}{-1}{0}\gsharm{s\p}{2}{0}\notag\\ 
&+2\om{2}{s\p}\om{0}{l}\overline{\gsharm{l\p}{-1}{m}}\gsharm{l}{0}{m}\gsharm{s}{-1}{0}\gsharm{s\p}{0}{0}
-2\om{0}{s\p}\om{0}{l}\overline{\gsharm{l\p}{1}{m}}\gsharm{l}{0}{m}\gsharm{s}{-1}{0}\gsharm{s\p}{2}{0}\notag\\
&+\om{2}{s\p}\om{2}{l}\overline{\gsharm{l\p}{1}{m}}\gsharm{l}{2}{m}\gsharm{s}{-1}{0}\gsharm{s\p}{0}{0}
+\om{2}{s\p}\om{2}{l}\overline{\gsharm{l\p}{-1}{m}}\gsharm{l}{-2}{m}\gsharm{s}{1}{0}\gsharm{s\p}{0}{0}\notag\\
&\left.-\om{0}{s\p}\om{2}{l}\overline{\gsharm{l\p}{1}{m}}\gsharm{l}{-2}{m}\gsharm{s}{1}{0}\gsharm{s\p}{2}{0}
+2\om{2}{s\p}\om{0}{l}\overline{\gsharm{l\p}{1}{m}}\gsharm{l}{0}{m}\gsharm{s}{1}{0}\gsharm{s\p}{0}{0}\right)\label{app:eq:kernel:fin:9}\\
\mathcal{T}_{10}=&-\frac{1}{4}\om{0}{s}\om{0}{s\p}\om{0}{l\p}\om{0}{l}\left(\overline{\gsharm{l\p}{-1}{m}}\gsharm{l}{1}{m}\gsharm{s}{-1}{0}\gsharm{s\p}{-1}{0}
+\overline{\gsharm{l\p}{-1}{m}}\gsharm{l}{-1}{m}\gsharm{s}{1}{0}\gsharm{s\p}{-1}{0}\right.\notag\\
&+\overline{\gsharm{l\p}{1}{m}}\gsharm{l}{1}{m}\gsharm{s}{1}{0}\gsharm{s\p}{-1}{0}
+\overline{\gsharm{l\p}{-1}{m}}\gsharm{l}{-1}{m}\gsharm{s}{-1}{0}\gsharm{s\p}{1}{0}\notag\\
&+\overline{\gsharm{l\p}{1}{m}}\gsharm{l}{1}{m}\gsharm{s}{-1}{0}\gsharm{s\p}{1}{0}
\left.+\overline{\gsharm{l\p}{1}{m}}\gsharm{l}{-1}{m}\gsharm{s}{1}{0}\gsharm{s\p}{1}{0}\right)\label{app:eq:kernel:fin:10}\\
\mathcal{T}_{11}=&\frac{1}{8}\om{0}{s}\om{0}{s\p}\om{0}{l\p}\om{0}{l}\left(
2\om{0}{l}\om{0}{l}\overline{\gsharm{l\p}{-1}{m}}\gsharm{l}{1}{m}\gsharm{s}{-1}{0}\gsharm{s\p}{-1}{0}\right.
+\om{2}{l}\om{2}{l}\overline{\gsharm{l\p}{-1}{m}}\gsharm{l}{1}{m}\gsharm{s}{-1}{0}\gsharm{s\p}{-1}{0}\notag\\
&+\om{2}{l}\om{3}{l}\overline{\gsharm{l\p}{1}{m}}\gsharm{l}{3}{m}\gsharm{s}{-1}{0}\gsharm{s\p}{-1}{0}		
+\om{2}{l}\om{2}{l}\overline{\gsharm{l\p}{-1}{m}}\gsharm{l}{-1}{m}\gsharm{s}{1}{0}\gsharm{s\p}{-1}{0}\notag\\
&+2\om{0}{l}\om{0}{l}\overline{\gsharm{l\p}{-1}{m}}\gsharm{l}{-1}{m}\gsharm{s}{1}{0}\gsharm{s\p}{-1}{0}
+2\om{0}{l}\om{0}{l}\overline{\gsharm{l\p}{1}{m}}\gsharm{l}{1}{m}\gsharm{s}{1}{0}\gsharm{s\p}{-1}{0}\notag\\
&+\om{2}{l}\om{2}{l}\overline{\gsharm{l\p}{1}{m}}\gsharm{l}{1}{m}\gsharm{s}{1}{0}\gsharm{s\p}{-1}{0}
+\om{2}{l}\om{2}{l}\overline{\gsharm{l\p}{-1}{m}}\gsharm{l}{-1}{m}\gsharm{s}{-1}{0}\gsharm{s\p}{1}{0}\notag\\
&+2\om{0}{l}\om{0}{l}\overline{\gsharm{l\p}{-1}{m}}\gsharm{l}{-1}{m}\gsharm{s}{-1}{0}\gsharm{s\p}{1}{0}
+2\om{0}{l}\om{0}{l}\overline{\gsharm{l\p}{1}{m}}\gsharm{l}{1}{m}\gsharm{s}{-1}{0}\gsharm{s\p}{1}{0}\notag\\
&+\om{2}{l}\om{2}{l}\overline{\gsharm{l\p}{1}{m}}\gsharm{l}{1}{m}\gsharm{s}{-1}{0}\gsharm{s\p}{1}{0}
+\om{2}{l}\om{3}{l}\overline{\gsharm{l\p}{-1}{m}}\gsharm{l}{-3}{m}\gsharm{s}{1}{0}\gsharm{s\p}{1}{0}\notag\\
&+\om{2}{l}\om{2}{l}\overline{\gsharm{l\p}{1}{m}}\gsharm{l}{-1}{m}\gsharm{s}{1}{0}\gsharm{s\p}{1}{0}
+\left.2\om{0}{l}\om{0}{l}\overline{\gsharm{l\p}{1}{m}}\gsharm{l}{-1}{m}\gsharm{s}{1}{0}\gsharm{s\p}{1}{0}\right)\label{app:eq:kernel:fin:11}\\
\mathcal{T}_{12} =&\frac{-m^2}{4}\om{0}{s}\om{0}{s\p}\om{0}{l}\om{0}{l\p}\left(4\om{0}{s}\om{0}{s}\om{0}{s'}\om{0}{s'}\overline{\gsharm{l\p}{-1}{m}}\gsharm{l}{-1}{m}\gsharm{s}{-1}{0}\gsharm{s'}{-1}{0}\gsharm{1}{1}{0}\gsharm{1}{1}{0}\notag\right.\\
&-2\om{0}{s}\om{0}{s}\om{2}{s'}\om{2}{s'}\overline{\gsharm{l\p}{-1}{m}}\gsharm{l}{-1}{m}\gsharm{s}{-1}{0}\gsharm{s'}{-1}{0}\gsharm{1}{1}{0}\gsharm{1}{1}{0}
-2\om{0}{s}\om{0}{s}\om{2}{s'}\om{3}{s'}\overline{\gsharm{l\p}{-1}{m}}\gsharm{l}{-1}{m}\gsharm{s}{1}{0}\gsharm{s'}{-3}{0}\gsharm{1}{1}{0}\gsharm{1}{1}{0}\notag\\
&-2\om{2}{s}\om{3}{s}\om{0}{s'}\om{0}{s'}\overline{\gsharm{l\p}{-1}{m}}\gsharm{l}{-1}{m}\gsharm{s}{-3}{0}\gsharm{s'}{1}{0}\gsharm{1}{1}{0}\gsharm{1}{1}{0} 
-2\om{2}{s}\om{2}{s}\om{0}{s'}\om{0}{s'}\overline{\gsharm{l\p}{-1}{m}}\gsharm{l}{-1}{m}\gsharm{s}{-1}{0}\gsharm{s'}{-1}{0}\gsharm{1}{1}{0}\gsharm{1}{1}{0}\notag\\
&+\om{2}{s}\om{2}{s}\om{2}{s'}\om{2}{s'}\overline{\gsharm{l\p}{-1}{m}}\gsharm{l}{-1}{m}\gsharm{s}{-1}{0}\gsharm{s'}{-1}{0}\gsharm{1}{1}{0}\gsharm{1}{1}{0}
-4\om{0}{s}\om{0}{s}\om{0}{s'}\overline{\gsharm{l\p}{-1}{m}}\gsharm{l}{-1}{m}\gsharm{s}{-1}{0}\gsharm{s'}{0}{0}\gsharm{1}{1}{0}\gsharm{1}{0}{0}\notag\\
&-2\om{0}{s}\om{0}{s}\om{2}{s'}\overline{\gsharm{l\p}{-1}{m}}\gsharm{l}{-1}{m}\gsharm{s}{1}{0}\gsharm{s'}{-2}{0}\gsharm{1}{1}{0}\gsharm{1}{0}{0}
+2\om{2}{s}\om{2}{s}\om{0}{s'}\overline{\gsharm{l\p}{-1}{m}}\gsharm{l}{-1}{m}\gsharm{s}{-1}{0}\gsharm{s'}{0}{0}\gsharm{1}{1}{0}\gsharm{1}{0}{0}\notag\\
&-2\om{2}{s}\om{0}{s'}\om{0}{s'}\overline{\gsharm{l\p}{-1}{m}}\gsharm{l}{-1}{m}\gsharm{s}{-2}{0}\gsharm{s'}{1}{0}\gsharm{1}{1}{0}\gsharm{1}{0}{0}
-4\om{0}{s}\om{0}{s'}\om{0}{s'}\overline{\gsharm{l\p}{-1}{m}}\gsharm{l}{-1}{m}\gsharm{s}{0}{0}\gsharm{s'}{-1}{0}\gsharm{1}{1}{0}\gsharm{1}{0}{0}\notag\\
&+2\om{0}{s}\om{2}{s'}\om{2}{s'}\overline{\gsharm{l\p}{-1}{m}}\gsharm{l}{-1}{m}\gsharm{s}{0}{0}\gsharm{s'}{-1}{0}\gsharm{1}{1}{0}\gsharm{1}{0}{0}
+\om{0}{s}\om{0}{s'}\gsharm{s}{0}{0}\overline{\gsharm{l\p}{-1}{m}}\gsharm{l}{-1}{m}\gsharm{s'}{0}{0}\gsharm{1}{0}{0}\gsharm{1}{0}{0}\notag\\
&+\om{2}{s}\om{2}{s}\om{2}{s'}\om{3}{s'}\overline{\gsharm{l\p}{-1}{m}}\gsharm{l}{-1}{m}\gsharm{s}{1}{0}\gsharm{s'}{-3}{0}\gsharm{1}{1}{0}\gsharm{1}{1}{0}
+\om{2}{s}\om{2}{s}\om{2}{s'}\gsharm{s}{1}{0}\overline{\gsharm{l\p}{-1}{m}}\gsharm{l}{-1}{m}\gsharm{s'}{-2}{0}\gsharm{1}{1}{0}\gsharm{1}{0}{0}\notag\\
&+\om{2}{s}\om{2}{s'}\om{3}{s'}\gsharm{s}{2}{0}\overline{\gsharm{l\p}{-1}{m}}\gsharm{l}{-1}{m}\gsharm{s'}{-3}{0}\gsharm{1}{1}{0}\gsharm{1}{0}{0}
+\om{2}{s}\om{2}{s'}\overline{\gsharm{l\p}{-1}{m}}\gsharm{l}{-1}{m}\gsharm{s}{2}{0}\gsharm{s'}{-2}{0}\gsharm{1}{0}{0}\gsharm{1}{0}{0}\notag\\
&+\om{2}{s}\om{3}{s}\om{2}{s'}\om{2}{s'}\overline{\gsharm{l\p}{-1}{m}}\gsharm{l}{-1}{m}\gsharm{s}{-3}{0}\gsharm{s'}{1}{0}\gsharm{1}{1}{0}\gsharm{1}{1}{0}
+\om{2}{s}\om{3}{s}\om{2}{s'}\overline{\gsharm{l\p}{-1}{m}}\gsharm{l}{-1}{m}\gsharm{s}{-3}{0}\gsharm{s'}{2}{0}\gsharm{1}{1}{0}\gsharm{1}{0}{0}\notag\\
&+\om{2}{s}\om{2}{s'}\om{2}{s'}\overline{\gsharm{l\p}{-1}{m}}\gsharm{l}{-1}{m}\gsharm{s}{-2}{0}\gsharm{s'}{1}{0}\gsharm{1}{1}{0}\gsharm{1}{0}{0}
+\om{0}{s}\om{0}{s'}\overline{\gsharm{l\p}{-1}{m}}\gsharm{l}{-1}{m}\gsharm{s}{0}{0}\gsharm{s'}{0}{0}\gsharm{1}{0}{0}\gsharm{1}{0}{0}\notag\\
&+4\om{0}{s}\om{0}{s}\om{0}{s'}\om{0}{s'}\overline{\gsharm{l\p}{1}{m}}\gsharm{l}{-1}{m}\gsharm{s}{-1}{0}\gsharm{s'}{1}{0}\gsharm{1}{1}{0}\gsharm{1}{1}{0}
+4\om{0}{s}\om{0}{s}\om{0}{s'}\om{0}{s'}\overline{\gsharm{l\p}{1}{m}}\gsharm{l}{-1}{m}\gsharm{s}{1}{0}\gsharm{s'}{-1}{0}\gsharm{1}{1}{0} \gsharm{1}{1}{0}\notag\\
&-2\om{0}{s}\om{0}{s}\om{2}{s'}\om{2}{s'}\overline{\gsharm{l\p}{1}{m}}\gsharm{l}{-1}{m}\gsharm{s}{1}{0}\gsharm{s'}{-1}{0}\gsharm{1}{1}{0} \gsharm{1}{1}{0}
-2\om{2}{s}\om{2}{s}\om{0}{s'}\om{0}{s'}\overline{\gsharm{l\p}{1}{m}}\gsharm{l}{-1}{m}\gsharm{s}{-1}{0}\gsharm{s'}{1}{0}\gsharm{1}{1}{0}\gsharm{1}{1}{0}\notag\\
&-4\om{0}{s}\om{0}{s}\om{0}{s'}\overline{\gsharm{l\p}{1}{m}}\gsharm{l}{-1}{m}\gsharm{s}{1}{0}\gsharm{s'}{0}{0}\gsharm{1}{1}{0}\gsharm{1}{0}{0}
-4\om{0}{s}\om{0}{s'}\om{0}{s'}\overline{\gsharm{l\p}{1}{m}}\gsharm{l}{-1}{m}\gsharm{s}{0}{0}\gsharm{s'}{1}{0}\gsharm{1}{1}{0}\gsharm{1}{0}{0}\notag\\
&-2\om{2}{s}\om{2}{s}\om{0}{s'}\om{0}{s'}\overline{\gsharm{l\p}{1}{m}}\gsharm{l}{-1}{m}\gsharm{s}{1}{0}\gsharm{s'}{-1}{0}\gsharm{1}{1}{0}\gsharm{1}{1}{0}
+\om{2}{s}\om{2}{s}\om{2}{s'}\om{2}{s'}\overline{\gsharm{l\p}{1}{m}}\gsharm{l}{-1}{m}\gsharm{s}{1}{0}\gsharm{s'}{-1}{0}\gsharm{1}{1}{0}\gsharm{1}{1}{0}\notag\\
&+\om{2}{s}\om{3}{s}\om{2}{s'}\om{3}{s'}\overline{\gsharm{l\p}{1}{m}}\gsharm{l}{-1}{m}\gsharm{s}{3}{0}\gsharm{s'}{-3}{0}\gsharm{1}{1}{0} \gsharm{1}{1}{0}
+2\om{2}{s}\om{2}{s}\om{0}{s'}\overline{\gsharm{l\p}{1}{m}}\gsharm{l}{-1}{m}\gsharm{s}{1}{0}\gsharm{s'}{0}{0}\gsharm{1}{1}{0}\gsharm{1}{0}{0}\notag\\
&+\om{2}{s}\om{3}{s}\om{2}{s'}\overline{\gsharm{l\p}{1}{m}}\gsharm{l}{-1}{m}\gsharm{s}{3}{0}\gsharm{s'}{-2}{0}\gsharm{1}{1}{0}\gsharm{1}{0}{0}
-2\om{2}{s}\om{0}{s'}\om{0}{s'}\overline{\gsharm{l\p}{1}{m}}\gsharm{l}{-1}{m}\gsharm{s}{2}{0}\gsharm{s'}{-1}{0}\gsharm{1}{1}{0}\gsharm{1}{0}{0}\notag\\
&+\om{2}{s}\om{2}{s'}\om{2}{s'}\overline{\gsharm{l\p}{1}{m}}\gsharm{l}{-1}{m}\gsharm{s}{2}{0}\gsharm{s'}{-1}{0}\gsharm{1}{1}{0}\gsharm{1}{0}{0}
+2\om{2}{s}\om{0}{s'}\gsharm{s}{2}{0}\overline{\gsharm{l\p}{1}{m}}\gsharm{l}{-1}{m}\gsharm{s'}{0}{0}\gsharm{1}{0}{0}\gsharm{1}{0}{0}\notag\\
&-2\om{0}{s}\om{0}{s}\om{2}{s'}\om{2}{s'}\overline{\gsharm{l\p}{1}{m}}\gsharm{l}{-1}{m}\gsharm{s}{-1}{0}\gsharm{s'}{1}{0}\gsharm{1}{1}{0}\gsharm{1}{1}{0}
+\om{2}{s}\om{3}{s}\om{2}{s'}\om{3}{s'}\overline{\gsharm{l\p}{1}{m}}\gsharm{l}{-1}{m}\gsharm{s}{-3}{0}\gsharm{s'}{3}{0}\gsharm{1}{1}{0}\gsharm{1}{1}{0}\notag\\
&+\om{2}{s}\om{2}{s}\om{2}{s'}\om{2}{s'}\overline{\gsharm{l\p}{1}{m}}\gsharm{l}{-1}{m}\gsharm{s}{-1}{0}\gsharm{s'}{1}{0}\gsharm{1}{1}{0}\gsharm{1}{1}{0}
-2\om{0}{s}\om{0}{s}\om{2}{s'}\overline{\gsharm{l\p}{1}{m}}\gsharm{l}{-1}{m}\gsharm{s}{-1}{0}\gsharm{s'}{2}{0}\gsharm{1}{1}{0}\gsharm{1}{0}{0}\notag\\
&+\om{2}{s}\om{2}{s}\om{2}{s'}\overline{\gsharm{l\p}{1}{m}}\gsharm{l}{-1}{m}\gsharm{s}{-1}{0}\gsharm{s'}{2}{0}\gsharm{1}{1}{0}\gsharm{1}{0}{0}
+\om{2}{s}\om{2}{s'}\om{3}{s'}\overline{\gsharm{l\p}{1}{m}}\gsharm{l}{-1}{m}\gsharm{s}{-2}{0}\gsharm{s'}{3}{0}\gsharm{1}{1}{0}\gsharm{1}{0}{0}\notag\\
&+2\om{0}{s}\om{2}{s'}\om{2}{s'}\overline{\gsharm{l\p}{1}{m}}\gsharm{l}{-1}{m}\gsharm{s}{0}{0}\gsharm{s'}{1}{0}\gsharm{1}{1}{0}\gsharm{1}{0}{0}
+2\om{0}{s}\om{2}{s'}\gsharm{s}{0}{0}\overline{\gsharm{l\p}{1}{m}}\gsharm{l}{-1}{m}\gsharm{s'}{2}{0}\gsharm{1}{0}{0}\gsharm{1}{0}{0}\notag\\
&-2\om{0}{s}\om{0}{s}\om{2}{s'}\om{3}{s'}\overline{\gsharm{l\p}{-1}{m}}\gsharm{l}{1}{m}\gsharm{s}{-1}{0}\gsharm{s'}{-3}{0}\gsharm{1}{1}{0}\gsharm{1}{1}{0}
-2\om{2}{s}\om{3}{s}\om{0}{s'}\om{0}{s'}\overline{\gsharm{l\p}{-1}{m}}\gsharm{l}{1}{m}\gsharm{s}{-3}{0}\gsharm{s'}{-1}{0}\gsharm{1}{1}{0}\gsharm{1}{1}{0}\notag\\
&+\om{2}{s}\om{3}{s}\om{2}{s'}\om{2}{s'}\overline{\gsharm{l\p}{-1}{m}}\gsharm{l}{1}{m}\gsharm{s}{-3}{0}\gsharm{s'}{-1}{0}\gsharm{1}{1}{0}\gsharm{1}{1}{0}
+\om{2}{s}\om{2}{s}\om{2}{s'}\om{3}{s'}\overline{\gsharm{l\p}{-1}{m}}\gsharm{l}{1}{m}\gsharm{s}{-1}{0}\gsharm{s'}{-3}{0}\gsharm{1}{1}{0}\gsharm{1}{1}{0}\notag\\
&-2\om{0}{s}\om{0}{s}\om{2}{s'}\overline{\gsharm{l\p}{-1}{m}}\gsharm{l}{1}{m}\gsharm{s}{-1}{0}\gsharm{s'}{-2}{0}\gsharm{1}{1}{0}\gsharm{1}{0}{0}
+2\om{2}{s}\om{3}{s}\om{0}{s'}\overline{\gsharm{l\p}{-1}{m}}\gsharm{l}{1}{m}\gsharm{s}{-3}{0}\gsharm{s'}{0}{0}\gsharm{1}{1}{0}\gsharm{1}{0}{0}\notag\\
&+\om{2}{s}\om{2}{s}\om{2}{s'}\overline{\gsharm{l\p}{-1}{m}}\gsharm{l}{1}{m}\gsharm{s}{-1}{0}\gsharm{s'}{-2}{0}\gsharm{1}{1}{0}\gsharm{1}{0}{0}
-2\om{2}{s}\om{0}{s'}\om{0}{s'}\overline{\gsharm{l\p}{-1}{m}}\gsharm{l}{1}{m}\gsharm{s}{-2}{0}\gsharm{s'}{-1}{0}\gsharm{1}{1}{0}\gsharm{1}{0}{0}\notag\\
&+\om{2}{s}\om{2}{s'}\om{2}{s'}\overline{\gsharm{l\p}{-1}{m}}\gsharm{l}{1}{m}\gsharm{s}{-2}{0}\gsharm{s'}{-1}{0}\gsharm{1}{1}{0}\gsharm{1}{0}{0}
+2\om{0}{s}\om{2}{s'}\om{3}{s'}\overline{\gsharm{l\p}{-1}{m}}\gsharm{l}{1}{m}\gsharm{s}{0}{0}\gsharm{s'}{-3}{0}\gsharm{1}{1}{0}\gsharm{1}{0}{0}\notag\\
&+2\om{2}{s}\om{0}{s'}\overline{\gsharm{l\p}{-1}{m}}\gsharm{l}{1}{m}\gsharm{s}{-2}{0}\gsharm{s'}{0}{0}\gsharm{1}{0}{0}\gsharm{1}{0}{0}
+2\om{0}{s}\om{2}{s'}\overline{\gsharm{l\p}{-1}{m}}\gsharm{l}{1}{m}\gsharm{s}{0}{0}\gsharm{s'}{-2}{0}\gsharm{1}{0}{0}\gsharm{1}{0}{0}\notag\\
&+4\om{0}{s}\om{0}{s}\om{0}{s'}\om{0}{s'}\overline{\gsharm{l\p}{1}{m}}\gsharm{l}{1}{m}\gsharm{s}{-1}{0}\gsharm{s'}{-1}{0}\gsharm{1}{1}{0}\gsharm{1}{1}{0}
-2\om{0}{s}\om{0}{s}\om{2}{s'}\om{2}{s'}\overline{\gsharm{l\p}{1}{m}}\gsharm{l}{1}{m}\gsharm{s}{-1}{0}\gsharm{s'}{-1}{0}\gsharm{1}{1}{0}\gsharm{1}{1}{0}\notag\\
&-2\om{0}{s}\om{0}{s}\om{2}{s'}\om{3}{s'}\overline{\gsharm{l\p}{1}{m}}\gsharm{l}{1}{m}\gsharm{s}{1}{0}\gsharm{s'}{-3}{0}\gsharm{1}{1}{0} \gsharm{1}{1}{0}
-2\om{2}{s}\om{3}{s}\om{0}{s'}\om{0}{s'}\overline{\gsharm{l\p}{1}{m}}\gsharm{l}{1}{m}\gsharm{s}{-3}{0}\gsharm{s'}{1}{0}\gsharm{1}{1}{0}\gsharm{1}{1}{0}\notag\\
&-2\om{2}{s}\om{2}{s}\om{0}{s'}\om{0}{s'}\overline{\gsharm{l\p}{1}{m}}\gsharm{l}{1}{m}\gsharm{s}{-1}{0}\gsharm{s'}{-1}{0}\gsharm{1}{1}{0}\gsharm{1}{1}{0}
+\om{2}{s}\om{2}{s}\om{2}{s'}\om{2}{s'}\overline{\gsharm{l\p}{1}{m}}\gsharm{l}{1}{m}\gsharm{s}{-1}{0}\gsharm{s'}{-1}{0}\gsharm{1}{1}{0}\gsharm{1}{1}{0}\notag\\
&-2\om{0}{s}\om{0}{s}\om{0}{s'}\overline{\gsharm{l\p}{1}{m}}\gsharm{l}{1}{m}\gsharm{s}{-1}{0}\gsharm{s'}{0}{0}\gsharm{1}{1}{0}\gsharm{1}{0}{0}
-2\om{0}{s}\om{0}{s}\om{2}{s'}\overline{\gsharm{l\p}{1}{m}}\gsharm{l}{1}{m}\gsharm{s}{1}{0}\gsharm{s'}{-2}{0}\gsharm{1}{1}{0}\gsharm{1}{0}{0}\notag\\
&+2\om{2}{s}\om{2}{s}\om{0}{s'}\overline{\gsharm{l\p}{1}{m}}\gsharm{l}{1}{m}\gsharm{s}{-1}{0}\gsharm{s'}{0}{0}\gsharm{1}{1}{0}\gsharm{1}{0}{0}
-2\om{2}{s}\om{0}{s'}\om{0}{s'}\overline{\gsharm{l\p}{1}{m}}\gsharm{l}{1}{m}\gsharm{s}{-2}{0}\gsharm{s'}{1}{0}\gsharm{1}{1}{0}\gsharm{1}{0}{0}\notag\\
&-4\om{0}{s}\om{0}{s'}\om{0}{s'}\overline{\gsharm{l\p}{1}{m}}\gsharm{l}{1}{m}\gsharm{s}{0}{0}\gsharm{s'}{-1}{0}\gsharm{1}{1}{0}\gsharm{1}{0}{0}
+2\om{0}{s}\om{2}{s'}\om{2}{s'}\overline{\gsharm{l\p}{1}{m}}\gsharm{l}{1}{m}\gsharm{s}{0}{0}\gsharm{s'}{-1}{0}\gsharm{1}{1}{0}\gsharm{1}{0}{0}\notag\\
&+4\om{0}{s}\om{0}{s'}\overline{\gsharm{l\p}{1}{m}}\gsharm{l}{1}{m}\gsharm{s}{0}{0}\gsharm{s'}{0}{0}\gsharm{1}{0}{0}\gsharm{1}{0}{0}
+\om{2}{s}\om{2}{s}\om{2}{s'}\om{3}{s'}\overline{\gsharm{l\p}{1}{m}}\gsharm{l}{1}{m}\gsharm{s}{1}{0}\gsharm{s'}{-3}{0}\gsharm{1}{1}{0}\gsharm{1}{1}{0}\notag\\
&+\om{2}{s}\om{2}{s}\om{2}{s'}\overline{\gsharm{l\p}{1}{m}}\gsharm{l}{1}{m}\gsharm{s}{1}{0}\gsharm{s'}{-2}{0}\gsharm{1}{1}{0}\gsharm{1}{0}{0}
+\om{2}{s}\om{2}{s'}\om{3}{s'}\overline{\gsharm{l\p}{1}{m}}\gsharm{l}{1}{m}\gsharm{s}{2}{0}\gsharm{s'}{-3}{0}\gsharm{1}{1}{0}\gsharm{1}{0}{0}\notag\\
&+2\om{2}{s}\om{2}{s'}\overline{\gsharm{l\p}{1}{m}}\gsharm{l}{1}{m}\gsharm{s}{2}{0}\gsharm{s'}{-2}{0}\gsharm{1}{0}{0}\gsharm{1}{0}{0}
+\om{2}{s}\om{3}{s}\om{2}{s'}\om{2}{s'}\overline{\gsharm{l\p}{1}{m}}\gsharm{l}{1}{m}\gsharm{s}{-3}{0}\gsharm{s'}{1}{0}\gsharm{1}{1}{0}\gsharm{1}{1}{0}\notag\\
&+\om{2}{s}\om{3}{s}\om{2}{s'}\overline{\gsharm{l\p}{1}{m}}\gsharm{l}{1}{m}\gsharm{s}{-3}{0}\gsharm{s'}{2}{0}\gsharm{1}{1}{0}\gsharm{1}{0}{0}
\left.+\om{2}{s}\om{2}{s'}\om{2}{s'}\overline{\gsharm{l\p}{1}{m}}\gsharm{l}{1}{m}\gsharm{s}{-2}{0}\gsharm{s'}{1}{0}\gsharm{1}{1}{0}\gsharm{1}{0}{0}\right)\label{app:eq:kernel:fin:12}\\
\mathcal{T}_{13}=&\frac{1}{8}\om{0}{s}\om{0}{s\p}\om{0}{l\p}\om{0}{l}\left(
2\om{0}{s}\om{0}{s}\overline{\gsharm{l\p}{-1}{m}}\gsharm{l}{-1}{m}\gsharm{s}{1}{0}\gsharm{s\p}{-1}{0}
+\om{2}{s}\om{2}{s}\overline{\gsharm{l\p}{-1}{m}}\gsharm{l}{-1}{m}\gsharm{s}{1}{0}\gsharm{s\p}{-1}{0}\right.\notag\\
&+\om{2}{s}\om{2}{s}\overline{\gsharm{l\p}{-1}{m}}\gsharm{l}{-1}{m}\gsharm{s}{-1}{0}\gsharm{s\p}{1}{0}
+2\om{0}{s}\om{0}{s}\overline{\gsharm{l\p}{-1}{m}}\gsharm{l}{-1}{m}\gsharm{s}{-1}{0}\gsharm{s\p}{1}{0}\notag\\
&+\om{2}{s}\om{3}{s}\overline{\gsharm{l\p}{1}{m}}\gsharm{l}{-1}{m}\gsharm{s}{3}{0}\gsharm{s\p}{-1}{0}
+2\om{0}{s}\om{0}{s}\overline{\gsharm{l\p}{1}{m}}\gsharm{l}{-1}{m}\gsharm{s}{1}{0}\gsharm{s\p}{1}{0}\notag\\
&+\om{2}{s}\om{2}{s}\overline{\gsharm{l\p}{1}{m}}\gsharm{l}{-1}{m}\gsharm{s}{1}{0}\gsharm{s\p}{1}{0}	
+\om{2}{s}\om{2}{s}\overline{\gsharm{l\p}{-1}{m}}\gsharm{l}{1}{m}\gsharm{s}{-1}{0}\gsharm{s\p}{-1}{0}\notag\\
&+2\om{0}{s}\om{0}{s}\overline{\gsharm{l\p}{-1}{m}}\gsharm{l}{1}{m}\gsharm{s}{-1}{0}\gsharm{s\p}{-1}{0}
+\om{2}{s}\om{3}{s}\overline{\gsharm{l\p}{-1}{m}}\gsharm{l}{1}{m}\gsharm{s}{-3}{0}\gsharm{s\p}{1}{0}\notag\\
&+2\om{0}{s}\om{0}{s}\overline{\gsharm{l\p}{1}{m}}\gsharm{l}{1}{m}\gsharm{s}{1}{0}\gsharm{s\p}{-1}{0}
+\om{2}{s}\om{2}{s}\overline{\gsharm{l\p}{1}{m}}\gsharm{l}{1}{m}\gsharm{s}{1}{0}\gsharm{s\p}{-1}{0}\notag\\
&+\om{2}{s}\om{2}{s}\overline{\gsharm{l\p}{1}{m}}\gsharm{l}{1}{m}\gsharm{s}{-1}{0}\gsharm{s\p}{1}{0}	
\left.+2\om{0}{s}\om{0}{s}\overline{\gsharm{l\p}{1}{m}}\gsharm{l}{1}{m}\gsharm{s}{-1}{0}\gsharm{s\p}{1}{0}\right)\label{app:eq:kernel:fin:13}\\
\mathcal{T}_{14}=&\frac{1}{4}\om{0}{s}\om{0}{s\p}\om{0}{l}\left(\om{2}{s\p}\overline{\gsharm{l\p}{0}{m}}\gsharm{l}{-1}{m}\gsharm{s}{-1}{0}\gsharm{s\p}{2}{0}
+2\om{0}{s\p}\overline{\gsharm{l\p}{0}{m}}\gsharm{l}{-1}{m}\gsharm{s}{1}{0}\gsharm{s\p}{0}{0}\right.\notag\\					  	&\left.+2\om{0}{s\p}\overline{\gsharm{l\p}{0}{m}}\gsharm{l}{1}{m}\gsharm{s}{-1}{0}\gsharm{s\p}{0}{0}
+\om{2}{s\p}\overline{\gsharm{l\p}{0}{m}}\gsharm{l}{1}{m}\gsharm{s}{1}{0}\gsharm{s\p}{-2}{0}\right)\label{app:eq:kernel:fin:14}\\
\mathcal{T}_{15}=&\frac{1}{8}\om{0}{s}\om{0}{s\p}\om{0}{l\p}\om{0}{l}\left(
4\om{0}{s}\om{0}{s\p}\overline{\gsharm{l\p}{-1}{m}}\gsharm{l}{-1}{m} \gsharm{s}{0}{0} \gsharm{s\p}{0}{0}
-\om{2}{s}\om{2}{s\p}\overline{\gsharm{l\p}{-1}{m}}\gsharm{l}{-1}{m} \gsharm{s}{-2}{0}\gsharm{s\p}{2}{0}\right.\notag\\
&-\om{2}{s}\om{2}{s\p}\overline{\gsharm{l\p}{-1}{m}}\gsharm{l}{-1}{m}\gsharm{s}{2}{0} \gsharm{s\p}{-2}{0}
-2\om{2}{s}\om{0}{s\p}\overline{\gsharm{l\p}{1}{m}}\gsharm{l}{-1}{m}  \gsharm{s}{2}{0} \gsharm{s\p}{0}{0}\notag\\
&+2\om{0}{s}\om{2}{s\p}\overline{\gsharm{l\p}{1}{m}}\gsharm{l}{-1}{m}  \gsharm{s}{0}{0}\gsharm{s\p}{2}{0}
-2\om{2}{s}\om{0}{s\p}\overline{\gsharm{l\p}{-1}{m}}\gsharm{l}{1}{m} \gsharm{s}{-2}{0}\gsharm{s\p}{0}{0}\notag\\
&+2\om{0}{s}\om{2}{s\p}\overline{\gsharm{l\p}{-1}{m}}\gsharm{l}{1}{m}\gsharm{s}{0}{0}\gsharm{s\p}{-2}{0}
-\om{2}{s}\om{2}{s\p}\overline{\gsharm{l\p}{1}{m}}\gsharm{l}{1}{m}   \gsharm{s}{-2}{0}\gsharm{s\p}{2}{0}\notag\\
&\left.-\om{2}{s}\om{2}{s\p}\overline{\gsharm{l\p}{1}{m}}\gsharm{l}{1}{m}  \gsharm{s}{2}{0}\gsharm{s\p}{-2}{0}
+4\om{0}{s}\om{0}{s\p}\overline{\gsharm{l\p}{1}{m}}\gsharm{l}{1}{m}\gsharm{s}{0}{0}\gsharm{s\p}{0}{0}\right)\label{app:eq:kernel:fin:15}\\
\mathcal{T}_{16}=&\frac{1}{8}\om{0}{s}\om{0}{s\p}\om{0}{l\p}\om{0}{l}\left(
\om{2}{s}\om{2}{s\p}\overline{\gsharm{l\p}{-1}{m}}\gsharm{l}{-1}{m}\gsharm{s}{2}{0}\gsharm{s\p}{-2}{0}
+4\om{0}{s}\om{0}{s\p}\overline{\gsharm{l\p}{-1}{m}}\gsharm{l}{-1}{m}\gsharm{s}{0}{0}\gsharm{s\p}{0}{0}\right.\notag\\
&+\om{2}{s}\om{2}{s\p}\overline{\gsharm{l\p}{-1}{m}}\gsharm{l}{-1}{m}\gsharm{s}{-2}{0}\gsharm{s\p}{2}{0}
+2\om{2}{s}\om{0}{s\p}\overline{\gsharm{l\p}{1}{m}}\gsharm{l}{-1}{m}\gsharm{s}{2}{0}\gsharm{s\p}{0}{0}\notag\\
&+2\om{0}{s}\om{2}{s\p}\overline{\gsharm{l\p}{1}{m}}\gsharm{l}{-1}{m}\gsharm{s}{0}{0}\gsharm{s\p}{2}{0}
+2\om{0}{s}\om{2}{s\p}\overline{\gsharm{l\p}{-1}{m}}\gsharm{l}{1}{m}\gsharm{s}{0}{0}\gsharm{s\p}{-2}{0}\notag\\
&+2\om{2}{s}\om{0}{s\p}\overline{\gsharm{l\p}{-1}{m}}\gsharm{l}{1}{m}\gsharm{s}{-2}{0}\gsharm{s\p}{0}{0}
+\om{2}{s}\om{2}{s\p}\overline{\gsharm{l\p}{1}{m}}\gsharm{l}{1}{m}\gsharm{s}{2}{0}\gsharm{s\p}{-2}{0}\notag\\
&\left.+4\om{0}{s}\om{0}{s\p}\overline{\gsharm{l\p}{1}{m}}\gsharm{l}{1}{m}\gsharm{s}{0}{0}\gsharm{s\p}{0}{0}
+\om{2}{s}\om{2}{s\p}\overline{\gsharm{l\p}{1}{m}}\gsharm{l}{1}{m}\gsharm{s}{-2}{0}\gsharm{s\p}{2}{0}\right)\label{app:eq:kernel:fin:16}\\
\mathcal{T}_{17}=&-\frac{m}{4}\sqrt{\frac{\pi}{3}}\om{0}{s}\om{0}{s\p}\om{0}{l}\om{0}{l\p}\notag\\
&\times\left(-2\om{2}{s}\om{0}{s\p}\om{0}{s\p}\overline{\gsharm{l\p}{-1}{m}}\gsharm{l}{-1}{m}\gsharm{s}{-2}{0}\gsharm{s\p}{1}{0}\gsharm{1}{1}{0}
-4\om{0}{s}\om{0}{s\p}\om{0}{s\p}\overline{\gsharm{l\p}{-1}{m}}\gsharm{l}{-1}{m}\gsharm{s}{0}{0}\gsharm{s\p}{-1}{0}\gsharm{1}{1}{0}\right.\notag\\
&+2\om{0}{s}\om{2}{s\p}\om{2}{s\p}\overline{\gsharm{l\p}{-1}{m}}\gsharm{l}{-1}{m}\gsharm{s}{0}{0}\gsharm{s\p}{-1}{0}\gsharm{1}{1}{0}
+4\om{0}{s}\om{0}{s\p}\overline{\gsharm{l\p}{-1}{m}}\gsharm{l}{-1}{m}\gsharm{s}{0}{0}\gsharm{s\p}{0}{0}\gsharm{1}{0}{0}\notag\\
&+4\om{0}{s}\om{0}{s\p}\om{0}{s\p}\overline{\gsharm{l\p}{1}{m}}\gsharm{l}{-1}{m}\gsharm{s}{0}{0}\gsharm{s\p}{1}{0}\gsharm{1}{1}{0}
+\om{2}{s}\om{2}{s\p}\om{3}{s\p}\overline{\gsharm{l\p}{-1}{m}}\gsharm{l}{-1}{m}\gsharm{s}{2}{0}\gsharm{s\p}{-3}{0}\gsharm{1}{1}{0}\notag\\
&+\om{2}{s}\om{2}{s\p}\overline{\gsharm{l\p}{-1}{m}}\gsharm{l}{-1}{m}\gsharm{s}{2}{0}\gsharm{s\p}{-2}{0}\gsharm{1}{0}{0}
+2\om{2}{s}\om{0}{s\p}\om{0}{s\p}\overline{\gsharm{l\p}{1}{m}}\gsharm{l}{-1}{m}\gsharm{s}{2}{0}\gsharm{s\p}{-1}{0}\gsharm{1}{1}{0}\notag\\
&-\om{2}{s}\om{2}{s\p}\om{2}{s\p}\overline{\gsharm{l\p}{1}{m}}\gsharm{l}{-1}{m}\gsharm{s}{2}{0}\gsharm{s\p}{-1}{0}\gsharm{1}{1}{0}
-2\om{2}{s}\om{0}{s\p}\overline{\gsharm{l\p}{1}{m}}\gsharm{l}{-1}{m}\gsharm{s}{2}{0}\gsharm{s\p}{0}{0}\gsharm{1}{0}{0}\notag\\
&+\om{2}{s}\om{2}{s\p}\om{2}{s\p}\overline{\gsharm{l\p}{-1}{m}}\gsharm{l}{-1}{m}\gsharm{s}{-2}{0}\gsharm{s\p}{1}{0}\gsharm{1}{1}{0}
+\om{2}{s}\om{2}{s\p}\overline{\gsharm{l\p}{-1}{m}}\gsharm{l}{-1}{m}\gsharm{s}{-2}{0}\gsharm{s\p}{2}{0}\gsharm{1}{0}{0}\notag\\
&-\om{2}{s}\om{2}{s\p}\om{3}{s\p}\overline{\gsharm{l\p}{1}{m}}\gsharm{l}{-1}{m}\gsharm{s}{-2}{0}\gsharm{s\p}{3}{0}\gsharm{1}{1}{0}
+\om{0}{s}\om{2}{s\p}\om{3}{s\p}\overline{\gsharm{l\p}{-1}{m}}\gsharm{l}{-1}{m}\gsharm{s}{0}{0}\gsharm{s\p}{-1}{0}\gsharm{1}{1}{0}\notag\\
&-2\om{0}{s}\om{2}{s\p}\om{2}{s\p}\overline{\gsharm{l\p}{1}{m}}\gsharm{l}{-1}{m}\gsharm{s}{0}{0}\gsharm{s\p}{1}{0}\gsharm{1}{1}{0}
-2\om{0}{s}\om{2}{s\p}\overline{\gsharm{l\p}{1}{m}}\gsharm{l}{-1}{m}\gsharm{s}{0}{0}\gsharm{s\p}{2}{0}\gsharm{1}{0}{0}\notag\\
&-2\om{2}{s}\om{0}{s\p}\om{0}{s\p}\overline{\gsharm{l\p}{-1}{m}}\gsharm{l}{1}{m}\gsharm{s}{-2}{0}\gsharm{s\p}{-1}{0}\gsharm{1}{1}{0}
+\om{2}{s}\om{2}{s\p}\om{2}{s\p}\overline{\gsharm{l\p}{-1}{m}}\gsharm{l}{1}{m}\gsharm{s}{-2}{0}\gsharm{s\p}{-1}{0}\gsharm{1}{1}{0}\notag\\
&+2\om{2}{s}\om{0}{s\p}\overline{\gsharm{l\p}{-1}{m}}\gsharm{l}{1}{m}\gsharm{s}{-2}{0}\gsharm{s\p}{0}{0}\gsharm{1}{0}{0}
+\om{2}{s}\om{0}{s\p}\om{0}{s\p}\overline{\gsharm{l\p}{1}{m}}\gsharm{l}{1}{m}\gsharm{s}{-2}{0}\gsharm{s\p}{1}{0}\gsharm{1}{1}{0}\notag\\
&+2\om{0}{s}\om{2}{s\p}\om{3}{s\p}\overline{\gsharm{l\p}{-1}{m}}\gsharm{l}{1}{m}\gsharm{s}{0}{0}\gsharm{s\p}{-3}{0}\gsharm{1}{1}{0}
+2\om{0}{s}\om{2}{s\p}\overline{\gsharm{l\p}{-1}{m}}\gsharm{l}{1}{m}\gsharm{s}{0}{0}\gsharm{s\p}{-2}{0}\gsharm{1}{0}{0}\notag\\
&+4\om{0}{s}\om{0}{s\p}\om{0}{s\p}\overline{\gsharm{l\p}{1}{m}}\gsharm{l}{1}{m}\gsharm{s}{0}{0}\gsharm{s\p}{-1}{0}\gsharm{1}{1}{0}
-2\om{0}{s}\om{2}{s\p}\om{2}{s\p}\overline{\gsharm{l\p}{1}{m}}\gsharm{l}{1}{m}\gsharm{s}{0}{0}\gsharm{s\p}{-1}{0}\gsharm{1}{1}{0}\notag\\
&-4\om{0}{s}\om{0}{s\p}\overline{\gsharm{l\p}{1}{m}}\gsharm{l}{1}{m}\gsharm{s}{0}{0}\gsharm{s\p}{0}{0}\gsharm{1}{0}{0}
-\om{2}{s}\om{2}{s\p}\om{3}{s\p}\overline{\gsharm{l\p}{1}{m}}\gsharm{l}{1}{m}\gsharm{s}{2}{0}\gsharm{s\p}{-3}{0}\gsharm{1}{1}{0}\notag\\		
&+\om{2}{s}\om{2}{s\p}\overline{\gsharm{l\p}{1}{m}}\gsharm{l}{1}{m}\gsharm{s}{2}{0}\gsharm{s\p}{-2}{0}\gsharm{1}{0}{0}
-\om{2}{s}\om{2}{s\p}\om{2}{s\p}\overline{\gsharm{l\p}{1}{m}}\gsharm{l}{1}{m}\gsharm{s}{-2}{0}\gsharm{s\p}{1}{0}\gsharm{1}{1}{0}\notag\\
&\left.+\om{2}{s}\om{0}{s\p}\om{0}{s\p}\overline{\gsharm{l\p}{1}{m}}\gsharm{l}{1}{m}\gsharm{s}{-2}{0}\gsharm{s\p}{1}{0}\gsharm{1}{1}{0}
-\om{2}{s}\om{2}{s\p}\overline{\gsharm{l\p}{1}{m}}\gsharm{l}{1}{m}\gsharm{s}{-2}{0}\gsharm{s\p}{2}{0}\gsharm{1}{0}{0}\right)\label{app:eq:kernel:fin:17}\\
\mathcal{T}_{18}=&-\frac{m}{4}\sqrt{\frac{\pi}{3}}\om{0}{s}\om{0}{s\p}\om{0}{l}\om{0}{l'}\notag\\
&\times\left(-4\om{0}{s}\om{0}{s'}\om{0}{s'}\overline{\gsharm{l'}{-1}{m}}\gsharm{l}{-1}{m} \gsharm{s}{0}{0}\gsharm{s'}{-1}{0}\gsharm{1}{1}{0}
+2\om{0}{s}\om{2}{s'}\om{2}{s'}\overline{\gsharm{l'}{-1}{m}}\gsharm{l}{-1}{m} \gsharm{s}{0}{0}\gsharm{s'}{-1}{0}\gsharm{1}{1}{0}\right.\notag\\
&+4\om{0}{s}\om{0}{s'}\overline{\gsharm{l'}{-1}{m}}\gsharm{l}{-1}{m} \gsharm{s}{0}{0}\gsharm{s'}{0}{0}\gsharm{1}{0}{0}  
+4\om{0}{s}\om{0}{s'}\om{0}{s'}\overline{\gsharm{l'}{1}{m}}\gsharm{l}{-1}{m} \gsharm{s}{0}{0}\gsharm{s'}{1}{0}\gsharm{1}{1}{0}\notag\\
&+2\om{2}{s}\om{0}{s'}\om{0}{s'}\overline{\gsharm{l'}{-1}{m}}\gsharm{l}{-1}{m} \gsharm{s}{-2}{0}\gsharm{s'}{1}{0}\gsharm{1}{1}{0} 
-\om{2}{s}\om{2}{s'}\om{3}{s'}\overline{\gsharm{l'}{-1}{m}}\gsharm{l}{-1}{m}\gsharm{s}{2}{0}\gsharm{s'}{-3}{0}\gsharm{1}{1}{0}\notag\\
&-2\om{2}{s}\om{2}{s'}\overline{\gsharm{l'}{-1}{m}}\gsharm{l}{-1}{m}\gsharm{s}{2}{0}\gsharm{s'}{-2}{0}\gsharm{1}{0}{0} -2\om{2}{s}\om{0}{s'}\om{0}{s'}\overline{\gsharm{l'}{1}{m}}\gsharm{l}{-1}{m}\gsharm{s}{2}{0}\gsharm{s'}{-1}{0}\gsharm{1}{1}{0}\notag\\
&+\om{2}{s}\om{2}{s'}\om{2}{s'}\overline{\gsharm{l'}{1}{m}}\gsharm{l}{-1}{m}\gsharm{s}{2}{0}\gsharm{s'}{-1}{0}\gsharm{1}{1}{0}
+2\om{2}{s}\om{0}{s'}\overline{\gsharm{l'}{1}{m}}\gsharm{l}{-1}{m}\gsharm{s}{2}{0}\gsharm{s'}{0}{0}\gsharm{1}{0}{0}\notag\\
&-\om{2}{s}\om{2}{s'}\om{2}{s'}\overline{\gsharm{l'}{-1}{m}}\gsharm{l}{-1}{m} \gsharm{s}{-2}{0}\gsharm{s'}{1}{0}\gsharm{1}{1}{0}
+\om{2}{s}\om{2}{s'}\om{3}{s'}\overline{\gsharm{l'}{1}{m}}\gsharm{l}{-1}{m} \gsharm{s}{-2}{0}\gsharm{s'}{3}{0}\gsharm{1}{1}{0}\notag\\
&-2\om{0}{s}\om{2}{s'}\om{2}{s'}\overline{\gsharm{l'}{1}{m}}\gsharm{l}{-1}{m}\gsharm{s}{0}{0}\gsharm{s'}{1}{0}\gsharm{1}{1}{0}
-2\om{0}{s}\om{2}{s'}\overline{\gsharm{l'}{1}{m}}\gsharm{l}{-1}{m}\gsharm{s}{0}{0}\gsharm{s'}{2}{0}\gsharm{1}{0}{0}\notag\\
&+2\om{0}{s}\om{2}{s'}\om{3}{s'}\overline{\gsharm{l'}{-1}{m}}\gsharm{l}{1}{m} \gsharm{s}{0}{0}\gsharm{s'}{-3}{0}\gsharm{1}{1}{0}
+2\om{0}{s}\om{2}{s'}\overline{\gsharm{l'}{-1}{m}}\gsharm{l}{1}{m} \gsharm{s}{0}{0}\gsharm{s'}{-2}{0}\gsharm{1}{0}{0}\notag\\
&+2\om{0}{s}\om{0}{s'}\om{0}{s'}\overline{\gsharm{l'}{1}{m}}\gsharm{l}{1}{m}\gsharm{s}{0}{0}\gsharm{s'}{-1}{0}\gsharm{1}{1}{0}
-2\om{0}{s}\om{2}{s'}\om{2}{s'}\overline{\gsharm{l'}{1}{m}}\gsharm{l}{1}{m}\gsharm{s}{0}{0}\gsharm{s'}{-1}{0}\gsharm{1}{1}{0}\notag\\
&+\om{2}{s}\om{0}{s'}\om{0}{s'}\overline{\gsharm{l'}{-1}{m}}\gsharm{l}{1}{m}\gsharm{s}{-2}{0}\gsharm{s'}{-1}{0}\gsharm{1}{1}{0} -\om{2}{s}\om{2}{s'}\om{2}{s'}\overline{\gsharm{l'}{-1}{m}}\gsharm{l}{1}{m} \gsharm{s}{-2}{0}\gsharm{s'}{-1}{0}\gsharm{1}{1}{0}\notag\\ 
&-\om{2}{s}\om{0}{s'}\om{0}{s'}\overline{\gsharm{l'}{1}{m}}\gsharm{l}{1}{m} \gsharm{s}{-2}{0}\gsharm{s'}{1}{0}\gsharm{1}{1}{0}
+\om{2}{s}\om{2}{s'}\om{3}{s'}\overline{\gsharm{l'}{1}{m}}\gsharm{l}{1}{m}\gsharm{s}{2}{0}\gsharm{s'}{-3}{0}\gsharm{1}{1}{0}\notag\\
&+\om{2}{s}\om{2}{s'}\overline{\gsharm{l'}{1}{m}}\gsharm{l}{1}{m}\gsharm{s}{2}{0}\gsharm{s'}{-2}{0}\gsharm{1}{0}{0}
+\om{2}{s}\om{0}{s'}\om{0}{s'}\overline{\gsharm{l'}{-1}{m}}\gsharm{l}{1}{m}\gsharm{s}{-2}{0}\gsharm{s'}{-1}{0}\gsharm{1}{1}{0}\notag\\
&-2\om{2}{s}\om{0}{s'}\overline{\gsharm{l'}{-1}{m}}\gsharm{l}{1}{m}\gsharm{s}{-2}{0}\gsharm{s'}{0}{0}\gsharm{1}{0}{0}
+\om{2}{s}\om{2}{s'}\om{2}{s'}\overline{\gsharm{l'}{1}{m}}\gsharm{l}{1}{m}\gsharm{s}{-2}{0}\gsharm{s'}{1}{0}\gsharm{1}{1}{0}\notag\\
&-\om{2}{s}\om{0}{s'}\om{0}{s'}\overline{\gsharm{l'}{1}{m}}\gsharm{l}{1}{m}\gsharm{s}{-2}{0}\gsharm{s'}{1}{0}\gsharm{1}{1}{0}
+\om{2}{s}\om{2}{s'}\overline{\gsharm{l'}{1}{m}}\gsharm{l}{1}{m}\gsharm{s}{-2}{0}\gsharm{s'}{2}{0}\gsharm{1}{0}{0}\notag\\
&\left.+2\om{0}{s}\om{0}{s'}\om{0}{s'}\overline{\gsharm{l'}{1}{m}}\gsharm{l}{1}{m}\gsharm{s}{0}{0}\gsharm{s'}{-1}{0}\gsharm{1}{1}{0}
-4\om{0}{s}\om{0}{s'}\overline{\gsharm{l'}{1}{m}}\gsharm{l}{1}{m}\gsharm{s}{0}{0}\gsharm{s'}{0}{0}\gsharm{1}{0}{0}\right)\label{app:eq:kernel:fin:18}\\
\mathcal{T}_{19}=& -\mathcal{T}_{3}\label{app:eq:kernel:fin:19}\\
\mathcal{T}_{20}=&\frac{1}{4}\om{0}{s}\om{0}{s\p}\om{0}{l}\left(
-\om{2}{s}\overline{\gsharm{l\p}{0}{m}}\gsharm{l}{-1}{m}\gsharm{s}{2}{0}\gsharm{s\p}{-1}{0}\right.
+2\om{0}{s}\overline{\gsharm{l\p}{0}{m}}\gsharm{l}{-1}{m}\gsharm{s}{0}{0} \gsharm{s\p}{1}{0}\notag\\
&\left.+2\om{0}{s}\overline{\gsharm{l\p}{0}{m}}\gsharm{l}{1}{m}\gsharm{s}{0}{0}\gsharm{s\p}{-1}{0}
-\om{2}{s}\overline{\gsharm{l\p}{0}{m}}\gsharm{l}{1}{m} \gsharm{s}{-2}{0}\gsharm{s\p}{1}{0}\right)\label{app:eq:kernel:fin:20}\\
\mathcal{T}_{21}=&\frac{1}{8}\om{0}{s}\om{0}{s\p}\om{0}{l\p}\om{0}{l}\left(4\om{0}{s}\om{0}{s\p} \overline{\gsharm{l\p}{-1}{m}}\gsharm{l}{-1}{m}\gsharm{s}{0}{0}\gsharm{s\p}{0}{0}
+\om{2}{s}\om{2}{s\p} \overline{\gsharm{l\p}{-1}{m}}\gsharm{l}{-1}{m}\gsharm{s}{2}{0}\gsharm{s\p}{-2}{0}\right.\notag\\
&+\om{2}{s}\om{2}{s\p} \overline{\gsharm{l\p}{-1}{m}}\gsharm{l}{-1}{m}\gsharm{s}{-2}{0}\gsharm{s\p}{2}{0}
-2\om{2}{s}\om{0}{s\p} \overline{\gsharm{l\p}{1}{m}}\gsharm{l}{-1}{m}\gsharm{s}{2}{0}\gsharm{s\p}{0}{0}\notag\\
&-2\om{0}{s}\om{2}{s\p} \overline{\gsharm{l\p}{1}{m}}\gsharm{l}{-1}{m}\gsharm{s}{0}{0}\gsharm{s\p}{2}{0}
-2\om{0}{s}\om{2}{s\p} \overline{\gsharm{l\p}{-1}{m}}\gsharm{l}{1}{m}\gsharm{s}{0}{0}\gsharm{s\p}{-2}{0}\notag\\
&-2\om{2}{s}\om{0}{s\p} \overline{\gsharm{l\p}{-1}{m}}\gsharm{l}{1}{m}\gsharm{s}{-2}{0}\gsharm{s\p}{0}{0}
+\om{2}{s}\om{2}{s\p} \overline{\gsharm{l\p}{1}{m}}\gsharm{l}{1}{m}\gsharm{s}{2}{0}\gsharm{s\p}{-2}{0}\notag\\
&\left.+\om{2}{s}\om{2}{s\p} \overline{\gsharm{l\p}{1}{m}}\gsharm{l}{1}{m}\gsharm{s}{-2}{0}\gsharm{s\p}{2}{0}
+4\om{0}{s}\om{0}{s\p} \overline{\gsharm{l\p}{1}{m}}\gsharm{l}{1}{m}\gsharm{s}{0}{0}\gsharm{s\p}{0}{0}\right)\label{app:eq:kernel:fin:21}\\
\mathcal{T}_{22}=&\frac{1}{8}\om{0}{s}\om{0}{s\p}\om{0}{l}\om{0}{l\p}\left(2\om{0}{s}\om{2}{s\p}\overline{\gsharm{l\p}{-1}{m}}\gsharm{l}{1}{m}\gsharm{s}{0}{0}\gsharm{s\p}{-2}{0}
+2\om{2}{s}\om{0}{s\p}\overline{\gsharm{l\p}{-1}{m}}\gsharm{l}{1}{m} \gsharm{s}{-2}{0}\gsharm{s\p}{0}{0}\right.\notag\\
&+4\om{0}{s}\om{0}{s\p}\overline{\gsharm{l\p}{1}{m}}\gsharm{l}{1}{m}\gsharm{s}{0}{0}\gsharm{s\p}{0}{0}
+\om{2}{s}\om{2}{s\p}\overline{\gsharm{l\p}{1}{m}}\gsharm{l}{1}{m} \gsharm{s}{2}{0}\gsharm{s\p}{-2}{0}\notag\\ 
&+\om{2}{s}\om{2}{s\p}\overline{\gsharm{l\p}{1}{m}}\gsharm{l}{1}{m}\gsharm{s}{-2}{0}\gsharm{s\p}{2}{0}
+4\om{0}{s}\om{0}{s\p}\overline{\gsharm{l\p}{-1}{m}}\gsharm{l}{-1}{m}\gsharm{s}{0}{0}\gsharm{s\p}{0}{0}\notag\\
&+\om{2}{s}\om{2}{s\p}\overline{\gsharm{l\p}{-1}{m}}\gsharm{l}{-1}{m} \gsharm{s}{2}{0}\gsharm{s\p}{-2}{0}
+\om{2}{s}\om{2}{s\p}\overline{\gsharm{l\p}{-1}{m}}\gsharm{l}{-1}{m} \gsharm{s}{-2}{0}\gsharm{s\p}{2}{0}\notag\\
&\left.+2\om{2}{s}\om{0}{s\p}\overline{\gsharm{l\p}{1}{m}}\gsharm{l}{-1}{m}\gsharm{s}{2}{0}\gsharm{s\p}{0}{0}
+2\om{0}{s}\om{2}{s\p}\overline{\gsharm{l\p}{1}{m}}\gsharm{l}{-1}{m}\gsharm{s}{0}{0}\gsharm{s\p}{2}{0}\right)\label{app:eq:kernel:fin:22}\\
\mathcal{T}_{23}=&\frac{1}{8}\om{0}{s}\om{0}{s\p}\om{0}{l\p}\om{0}{l}\left(
-2\om{2}{s}\om{0}{s\p}\overline{\gsharm{l\p}{-1}{m}}\gsharm{l}{1}{m}\gsharm{s}{-2}{0}\gsharm{s\p}{0}{0}
+4\om{0}{s}\om{0}{s\p}\overline{\gsharm{l\p}{1}{m}}\gsharm{l}{1}{m}\gsharm{s}{0}{0}\gsharm{s\p}{0}{0}\right.\notag\\
&+2\om{0}{s}\om{2}{s\p}\overline{\gsharm{l\p}{-1}{m}}\gsharm{l}{1}{m}\gsharm{s}{0}{0}\gsharm{s\p}{-2}{0}
-\om{2}{s}\om{2}{s\p}\overline{\gsharm{l\p}{1}{m}}\gsharm{l}{1}{m}\gsharm{s}{2}{0}\gsharm{s\p}{-2}{0}\notag\\
&-\om{2}{s}\om{2}{s\p}\overline{\gsharm{l\p}{-1}{m}}\gsharm{l}{-1}{m}\gsharm{s}{2}{0}\gsharm{s\p}{-2}{0}
-\om{2}{s}\om{2}{s\p}\overline{\gsharm{l\p}{1}{m}}\gsharm{l}{1}{m}\gsharm{s}{-2}{0}\gsharm{s\p}{2}{0}\notag\\
&-\om{2}{s}\om{2}{s\p}\overline{\gsharm{l\p}{-1}{m}}\gsharm{l}{-1}{m}\gsharm{s}{-2}{0}\gsharm{s\p}{2}{0}
+2\om{0}{s}\om{2}{s\p}\overline{\gsharm{l\p}{1}{m}}\gsharm{l}{-1}{m}\gsharm{s}{0}{0}\gsharm{s\p}{2}{0}\notag\\
&\left.+4\om{0}{s}\om{0}{s\p}\overline{\gsharm{l\p}{-1}{m}}\gsharm{l}{-1}{m}\gsharm{s}{0}{0}\gsharm{s\p}{0}{0}
-2\om{2}{s}\om{0}{s\p}\overline{\gsharm{l\p}{1}{m}}\gsharm{l}{-1}{m}\gsharm{s}{2}{0}\gsharm{s\p}{0}{0}\right)\label{app:eq:kernel:fin:23}\\
\mathcal{T}_{24}=& -\frac{m}{4}\sqrt{\frac{\pi}{3}}\om{0}{s}\om{0}{s\p}\om{0}{l}\om{0}{l\p}\notag\\
&\times\left(2\om{0}{s}\om{2}{s'}\om{3}{s'}\overline{\gsharm{l\p}{-1}{m}}\gsharm{l}{1}{m}\gsharm{s}{0}{0}\gsharm{s'}{-3}{0}\gsharm{1}{1}{0}
+2\om{0}{s}\om{2}{s'}\overline{\gsharm{l\p}{-1}{m}}\gsharm{l}{1}{m}\gsharm{s}{0}{0}\gsharm{s'}{-2}{0}\gsharm{1}{0}{0}\right.\notag\\ &+4\om{0}{s}\om{0}{s'}\om{0}{s'}\overline{\gsharm{l\p}{-1}{m}}\gsharm{l}{-1}{m}\gsharm{s}{0}{0}\gsharm{s'}{-1}{0}\gsharm{1}{1}{0}
-2\om{0}{s}\om{2}{s'}\om{2}{s'}\overline{\gsharm{l\p}{-1}{m}}\gsharm{l}{-1}{m}\gsharm{s}{0}{0}\gsharm{s'}{-1}{0}\gsharm{1}{1}{0}\notag\\
&-4\om{0}{s}\om{0}{s'}\overline{\gsharm{l\p}{-1}{m}}\gsharm{l}{-1}{m}\gsharm{s}{0}{0}\gsharm{s'}{0}{0}\gsharm{1}{0}{0}
+2\om{2}{s}\om{0}{s'}\om{0}{s'}\overline{\gsharm{l\p}{-1}{m}}\gsharm{l}{1}{m}\gsharm{s}{-2}{0}\gsharm{s'}{-1}{0}\gsharm{1}{1}{0}\notag\\
&-\om{2}{s}\om{2}{s'}\om{2}{s'}\overline{\gsharm{l\p}{-1}{m}}\gsharm{l}{1}{m}\gsharm{s}{-2}{0}\gsharm{s'}{-1}{0}\gsharm{1}{1}{0}
-\om{2}{s}\om{0}{s'}\overline{\gsharm{l\p}{-1}{m}}\gsharm{l}{1}{m}\gsharm{s}{-2}{0}\gsharm{s'}{0}{0}\gsharm{1}{0}{0}\notag\\
&-\om{2}{s}\om{0}{s'}\om{0}{s'}\overline{\gsharm{l\p}{-1}{m}}\gsharm{l}{-1}{m}\gsharm{s}{-2}{0}\gsharm{s'}{1}{0}\gsharm{1}{1}{0}
+\om{2}{s}\om{2}{s'}\om{3}{s'}\overline{\gsharm{l\p}{-1}{m}}\gsharm{l}{-1}{m}\gsharm{s}{2}{0}\gsharm{s'}{-3}{0}\gsharm{1}{1}{0}\notag\\
&+\om{2}{s}\om{2}{s'}\overline{\gsharm{l\p}{-1}{m}}\gsharm{l}{-1}{m}\gsharm{s}{2}{0}\gsharm{s'}{-2}{0}\gsharm{1}{0}{0}
-\om{2}{s}\om{0}{s'}\overline{\gsharm{l\p}{-1}{m}} \gsharm{l}{1}{m}\gsharm{s}{-2}{0}\gsharm{s'}{0}{0}\gsharm{1}{0}{0}\notag\\
&+\om{2}{s}\om{2}{s'}\om{2}{s'}\overline{\gsharm{l\p}{-1}{m}} \gsharm{l}{-1}{m}\gsharm{s}{-2}{0}\gsharm{s'}{1}{0}\gsharm{1}{1}{0}
-\om{2}{s}\om{0}{s'}\om{0}{s'}\overline{\gsharm{l\p}{-1}{m}} \gsharm{l}{-1}{m}\gsharm{s}{-2}{0}\gsharm{s'}{1}{0}\gsharm{1}{1}{0}\notag\\
&+\om{2}{s}\om{2}{s'}\overline{\gsharm{l\p}{-1}{m}} \gsharm{l}{-1}{m}\gsharm{s}{-2}{0}\gsharm{s'}{2}{0}\gsharm{1}{0}{0}
+2\om{2}{s}\om{0}{s'}\om{0}{s'}\overline{\gsharm{l\p}{1}{m}} \gsharm{l}{1}{m}\gsharm{s}{-2}{0}\gsharm{s'}{1}{0}\gsharm{1}{1}{0}\notag\\
&-\om{2}{s}\om{2}{s'}\om{3}{s'}\overline{\gsharm{l\p}{1}{m}}\gsharm{l}{1}{m}\gsharm{s}{2}{0}\gsharm{s'}{-3}{0}\gsharm{1}{1}{0}
-\om{2}{s}\om{2}{s'}\overline{\gsharm{l\p}{1}{m}}\gsharm{l}{1}{m}\gsharm{s}{2}{0}\gsharm{s'}{-2}{0}\gsharm{1}{0}{0}\notag\\ &-2\om{2}{s}\om{0}{s'}\om{0}{s'}\overline{\gsharm{l\p}{1}{m}}\gsharm{l}{-1}{m}\gsharm{s}{2}{0}\gsharm{s'}{-1}{0}\gsharm{1}{1}{0}
+\om{2}{s}\om{2}{s'}\om{2}{s'}\overline{\gsharm{l\p}{1}{m}}\gsharm{l}{-1}{m}\gsharm{s}{2}{0}\gsharm{s'}{-1}{0}\gsharm{1}{1}{0}\notag\\
&+2\om{2}{s}\om{0}{s'}\overline{\gsharm{l\p}{1}{m}}\gsharm{l}{-1}{m}\gsharm{s}{2}{0}\gsharm{s'}{0}{0}\gsharm{1}{0}{0}
-4\om{0}{s}\om{0}{s'}\om{0}{s'}\overline{\gsharm{l\p}{1}{m}} \gsharm{l}{1}{m}\gsharm{s}{0}{0}\gsharm{s'}{-1}{0}\gsharm{1}{1}{0}\notag\\
&+2\om{0}{s}\om{2}{s'}\om{2}{s'}\overline{\gsharm{l\p}{1}{m}} \gsharm{l}{1}{m}\gsharm{s}{0}{0}\gsharm{s'}{-1}{0}\gsharm{1}{1}{0}
+4\om{0}{s}\om{0}{s'}\overline{\gsharm{l\p}{1}{m}} \gsharm{l}{1}{m}\gsharm{s}{0}{0}\gsharm{s'}{0}{0}\gsharm{1}{0}{0}\notag\\
&+4\om{0}{s}\om{0}{s'}\om{0}{s'}\overline{\gsharm{l\p}{1}{m}} \gsharm{l}{-1}{m}\gsharm{s}{0}{0}\gsharm{s'}{1}{0}\gsharm{1}{1}{0}
-\om{2}{s}\om{2}{s'}\om{2}{s'}\overline{\gsharm{l\p}{1}{m}} \gsharm{l}{1}{m}\gsharm{s}{-2}{0}\gsharm{s'}{1}{0}\gsharm{1}{1}{0}\notag\\
&-\om{2}{s}\om{2}{s'}\overline{\gsharm{l\p}{1}{m}} \gsharm{l}{1}{m}\gsharm{s}{-2}{0}\gsharm{s'}{2}{0}\gsharm{1}{0}{0}
+\om{2}{s}\om{2}{s'}\om{3}{s'}\overline{\gsharm{l\p}{1}{m}} \gsharm{l}{-1}{m}\gsharm{s}{-2}{0}\gsharm{s'}{3}{0}\gsharm{1}{1}{0}\notag\\
&-2\om{0}{s}\om{2}{s'}\om{2}{s'}\overline{\gsharm{l\p}{1}{m}}\gsharm{l}{-1}{m}\gsharm{s}{0}{0}\gsharm{s'}{1}{0}\gsharm{1}{1}{0}
\left.-\om{0}{s}\om{2}{s'}\overline{\gsharm{l\p}{1}{m}}\gsharm{l}{-1}{m}\gsharm{s}{0}{0}\gsharm{s'}{2}{0}\gsharm{1}{0}{0}\right)\label{app:eq:kernel:fin:24}\\
\mathcal{T}_{25}=&\frac{m}{4}\sqrt{\frac{\pi}{3}}\om{0}{s}\om{0}{s\p}\om{0}{l}\om{0}{l\p}\notag\\
&\times\left(2\om{2}{s}\om{0}{s'}\om{0}{s'}\overline{\gsharm{l\p}{-1}{m}}\gsharm{l}{1}{m}\gsharm{s}{-2}{0}\gsharm{s'}{-1}{0}\gsharm{1}{1}{0}
-\om{2}{s}\om{2}{s'}\om{2}{s'}\overline{\gsharm{l\p}{-1}{m}}\gsharm{l}{1}{m}\gsharm{s}{-2}{0}\gsharm{s'}{-1}{0}\gsharm{1}{1}{0}\right.\notag\\
&-2\om{2}{s}\om{0}{s'}\overline{\gsharm{l\p}{-1}{m}}\gsharm{l}{1}{m}\gsharm{s}{-2}{0}\gsharm{s'}{0}{0}\gsharm{1}{0}{0}
-2\om{2}{s}\om{0}{s'}\om{0}{s'}\overline{\gsharm{l\p}{-1}{m}}\gsharm{l}{-1}{m}\gsharm{s}{-2}{0}\gsharm{s'}{1}{0}\gsharm{1}{1}{0}\notag\\
&-2\om{0}{s}\om{2}{s'}\om{3}{s'}\overline{\gsharm{l\p}{-1}{m}}\gsharm{l}{1}{m}\gsharm{s}{0}{0}\gsharm{s'}{-3}{0}\gsharm{1}{1}{0}
-2\om{0}{s}\om{2}{s'}\overline{\gsharm{l\p}{-1}{m}}\gsharm{l}{1}{m}\gsharm{s}{0}{0}\gsharm{s'}{-2}{0}\gsharm{1}{0}{0}\notag\\
&-4\om{0}{s}\om{0}{s'}\om{0}{s'}\overline{\gsharm{l\p}{-1}{m}}\gsharm{l}{-1}{m}\gsharm{s}{0}{0}\gsharm{s'}{-1}{0}\gsharm{1}{1}{0}
+2\om{0}{s}\om{2}{s'}\om{2}{s'}\overline{\gsharm{l\p}{-1}{m}}\gsharm{l}{-1}{m}\gsharm{s}{0}{0}\gsharm{s'}{-1}{0}\gsharm{1}{1}{0}\notag\\
&+4\om{0}{s}\om{0}{s'}\overline{\gsharm{l\p}{-1}{m}}\gsharm{l}{-1}{m}\gsharm{s}{0}{0}\gsharm{s'}{0}{0}\gsharm{1}{0}{0}
+\om{2}{s}\om{2}{s'}\om{3}{s'}\overline{\gsharm{l\p}{-1}{m}}\gsharm{l}{-1}{m}\gsharm{s}{2}{0}\gsharm{s'}{-3}{0}\gsharm{1}{1}{0}\notag\\
&+\om{2}{s}\om{2}{s'}\overline{\gsharm{l\p}{-1}{m}}\gsharm{l}{-1}{m}\gsharm{s}{2}{0}\gsharm{s'}{-2}{0}\gsharm{1}{0}{0}
+\om{2}{s}\om{2}{s'}\om{2}{s'}\overline{\gsharm{l\p}{-1}{m}}\gsharm{l}{-1}{m}\gsharm{s}{-2}{0}\gsharm{s'}{1}{0}\gsharm{1}{1}{0}\notag\\
&+\om{2}{s}\om{2}{s'}\overline{\gsharm{l\p}{-1}{m}}\gsharm{l}{-1}{m}\gsharm{s}{-2}{0}\gsharm{s'}{2}{0}\gsharm{1}{0}{0}
+2\om{2}{s}\om{0}{s'}\om{0}{s'}\overline{\gsharm{l\p}{1}{m}}\gsharm{l}{1}{m}\gsharm{s}{-2}{0}\gsharm{s'}{1}{0}\gsharm{1}{1}{0}\notag\\
&+4\om{0}{s}\om{0}{s'}\om{0}{s'}\overline{\gsharm{l\p}{1}{m}}\gsharm{l}{1}{m}\gsharm{s}{0}{0}\gsharm{s'}{-1}{0}\gsharm{1}{1}{0}
-2\om{0}{s}\om{2}{s'}\om{2}{s'}\overline{\gsharm{l\p}{1}{m}}\gsharm{l}{1}{m}\gsharm{s}{0}{0}\gsharm{s'}{-1}{0}\gsharm{1}{1}{0}\notag\\
&-4\om{0}{s}\om{0}{s'}\overline{\gsharm{l\p}{1}{m}}\gsharm{l}{1}{m}\gsharm{s}{0}{0}\gsharm{s'}{0}{0}\gsharm{1}{0}{0}
-4\om{0}{s}\om{0}{s'}\om{0}{s'}\overline{\gsharm{l\p}{1}{m}}\gsharm{l}{-1}{m}\gsharm{s}{0}{0}\gsharm{s'}{1}{0}\gsharm{1}{1}{0}\notag\\
&-\om{2}{s}\om{2}{s'}\om{3}{s'}\overline{\gsharm{l\p}{1}{m}}\gsharm{l}{1}{m}\gsharm{s}{2}{0}\gsharm{s'}{-3}{0}\gsharm{1}{1}{0}
-\om{2}{s}\om{2}{s'}\overline{\gsharm{l\p}{1}{m}}\gsharm{l}{1}{m}\gsharm{s}{2}{0}\gsharm{s'}{-2}{0}\gsharm{1}{0}{0}\notag\\
&-2\om{2}{s}\om{0}{s'}\om{0}{s'}\overline{\gsharm{l\p}{1}{m}}\gsharm{l}{-1}{m}\gsharm{s}{2}{0}\gsharm{s'}{-1}{0}\gsharm{1}{1}{0}
+\om{2}{s}\om{2}{s'}\om{2}{s'}\overline{\gsharm{l\p}{1}{m}}\gsharm{l}{-1}{m}\gsharm{s}{2}{0}\gsharm{s'}{-1}{0}\gsharm{1}{1}{0}\notag\\
&+2\om{2}{s}\om{0}{s'}\overline{\gsharm{l\p}{1}{m}}\gsharm{l}{-1}{m}\gsharm{s}{2}{0}\gsharm{s'}{0}{0}\gsharm{1}{0}{0}
-\om{2}{s}\om{2}{s'}\om{2}{s'}\overline{\gsharm{l\p}{1}{m}}\gsharm{l}{1}{m}\gsharm{s}{-2}{0}\gsharm{s'}{1}{0}\gsharm{1}{1}{0}\notag\\
&-\om{2}{s}\om{2}{s'}\overline{\gsharm{l\p}{1}{m}}\gsharm{l}{1}{m}\gsharm{s}{-2}{0}\gsharm{s'}{2}{0}\gsharm{1}{0}{0}
+\om{2}{s}\om{2}{s'}\om{3}{s'}\overline{\gsharm{l\p}{1}{m}}\gsharm{l}{-1}{m}\gsharm{s}{-2}{0}\gsharm{s'}{3}{0}\gsharm{1}{1}{0}\notag\\
&\left.+2\om{0}{s}\om{2}{s'}\om{2}{s'}\overline{\gsharm{l\p}{1}{m}}\gsharm{l}{-1}{m}\gsharm{s}{0}{0}\gsharm{s'}{1}{0}\gsharm{1}{1}{0}
+2\om{0}{s}\om{2}{s'}\overline{\gsharm{l\p}{1}{m}}\gsharm{l}{-1}{m}\gsharm{s}{0}{0}\gsharm{s'}{2}{0}\gsharm{1}{0}{0}\right)\label{app:eq:kernel:fin:25}
\end{align}


\bibliographystyle{astronurl_ads}
\bibliography{C:/Users/kiefer/Desktop/PhD/References/library} 

\begin{thebibliography}{}

\bibitem[\protect\astroncite{Aerts et~al.}{2010}]{Aerts2010}
Aerts, C., Christensen-Dalsgaard, J., and Kurtz, D.~W.: 2010,
\newblock {\em {Asteroseismology}},
\newblock Springer, Dordrecht,
\newblock {\small[\href{http://adsabs.harvard.edu/abs/2010aste.book.....A
  https://books.google.de/books/about/Asteroseismology.html?id=N8pswDrdSyUC{\&}redir{\_}esc=y}{URL}]}

\bibitem[\protect\astroncite{Antia et~al.}{2000}]{Antia2000}
Antia, H.~M., Chitre, S.~M., and Thompson, M.~J.: 2000,
\newblock {\em Astronomy {\&} Astrophysics} {\bf 360}, 335,
\newblock {\small[\href{http://adsabs.harvard.edu/abs/2000A{\&}A...360..335A
  http://adsabs.harvard.edu/cgi-bin/nph-data{\_}query?bibcode=2000A{\&}A...360..335A{\&}link{\_}type=ARTICLE{\&}db{\_}key=AST{\&}high=}{URL}]}

\bibitem[\protect\astroncite{Baldner et~al.}{2009}]{Baldner2009}
Baldner, C.~S., Antia, H.~M., Basu, S., and Larson, T.~P.: 2009,
\newblock {\em The Astrophysical Journal} {\bf 705(2)}, 1704,
\newblock
  {\small[\href{http://stacks.iop.org/0004-637X/705/i=2/a=1704?key=crossref.ac75f2fc6dc8baa50b3b7c2d03b09dfd}{URL}]},
\newblock {\small[\href{http://dx.doi.org/10.1088/0004-637X/705/2/1704}{DOI}]}

\bibitem[\protect\astroncite{Beck et~al.}{2012}]{Beck2012}
Beck, P.~G., Montalban, J., Kallinger, T., {De Ridder}, J., Aerts, C.,
  Garc{\'{i}}a, R.~A., Hekker, S., Dupret, M.-A., Mosser, B., Eggenberger, P.,
  Stello, D., Elsworth, Y., Frandsen, S., Carrier, F., Hillen, M., Gruberbauer,
  M., Christensen-Dalsgaard, J., Miglio, A., Valentini, M., Bedding, T.~R.,
  Kjeldsen, H., Girouard, F.~R., Hall, J.~R., and Ibrahim, K.~A.: 2012,
\newblock {\em Nature} {\bf 481(7379)}, 55,
\newblock
  {\small[\href{http://adsabs.harvard.edu/abs/2012Natur.481...55B}{URL}]},
\newblock {\small[\href{http://dx.doi.org/10.1038/nature10612}{DOI}]}

\bibitem[\protect\astroncite{Braun and Fan}{1998}]{Braun1998}
Braun, D.~C. and Fan, Y.: 1998,
\newblock {\em The Astrophysical Journal} {\bf 508(1)}, L105,
\newblock
  {\small[\href{http://adsabs.harvard.edu/abs/1998ApJ...508L.105B}{URL}]},
\newblock {\small[\href{http://dx.doi.org/10.1086/311727}{DOI}]}

\bibitem[\protect\astroncite{Broomhall}{2017}]{Broomhall2017}
Broomhall, A.-M.: 2017,
\newblock {\em Solar Physics} {\bf 292}, 67,
\newblock
  {\small[\href{http://adsabs.harvard.edu/abs/2017SoPh..292...67B}{URL}]},
\newblock {\small[\href{http://dx.doi.org/10.1007/s11207-017-1068-5}{DOI}]}

\bibitem[\protect\astroncite{Broomhall et~al.}{2015}]{Broomhall2015}
Broomhall, A.-M., Pugh, C., and Nakariakov, V.: 2015,
\newblock {\em Advances in Space Research} {\bf 56(12)}, 2706,
\newblock
  {\small[\href{http://adsabs.harvard.edu/abs/2015AdSpR..56.2706B}{URL}]},
\newblock {\small[\href{http://dx.doi.org/10.1016/j.asr.2015.04.018}{DOI}]}

\bibitem[\protect\astroncite{Caligari et~al.}{1995}]{Caligari1995}
Caligari, P., Moreno-Insertis, F., and Sch{\"{u}}ssler, M.: 1995,
\newblock {\em The Astrophysical Journal} {\bf 441}, 886,
\newblock
  {\small[\href{http://adsabs.harvard.edu/abs/1995ApJ...441..886C}{URL}]},
\newblock {\small[\href{http://dx.doi.org/10.1086/175410}{DOI}]}

\bibitem[\protect\astroncite{Charbonneau}{2010}]{Charbonneau2010}
Charbonneau, P.: 2010,
\newblock {\em Living Reviews in Solar Physics} {\bf 7(3)}, 3,
\newblock {\small[\href{http://www.livingreviews.org/lrsp-2010-3}{URL}]},
\newblock {\small[\href{http://dx.doi.org/10.12942/lrsp-2010-3}{DOI}]}

\bibitem[\protect\astroncite{Charbonneau}{2013}]{Charbonneau2013}
Charbonneau, P.: 2013,
\newblock {\em {Solar and Stellar Dynamos}}, Vol.~39 of {\em Saas-Fee Advanced
  Courses},
\newblock Springer Berlin Heidelberg, Berlin, Heidelberg,
\newblock
  {\small[\href{http://adsabs.harvard.edu/abs/2013SAAS...39.....C}{URL}]},
\newblock {\small[\href{http://dx.doi.org/10.1007/978-3-642-32093-4}{DOI}]}

\bibitem[\protect\astroncite{Charbonneau}{2014}]{Charbonneau2014}
Charbonneau, P.: 2014,
\newblock {\em Annual Review of Astronomy and Astrophysics} {\bf 52(1)}, 251,
\newblock
  {\small[\href{http://adsabs.harvard.edu/abs/2014ARA{\%}26A..52..251C}{URL}]},
\newblock
  {\small[\href{http://dx.doi.org/10.1146/annurev-astro-081913-040012}{DOI}]}

\bibitem[\protect\astroncite{Choudhuri et~al.}{1995}]{Choudhuri1995}
Choudhuri, A.~R., Sch{\"{u}}ssler, M., and Dikpati, M.: 1995,
\newblock {\em Astronomy and Astrophysics} {\bf 303}, L29,
\newblock
  {\small[\href{http://adsabs.harvard.edu/abs/1995A{\%}26A...303L..29C}{URL}]}

\bibitem[\protect\astroncite{Christensen-Dalsgaard
  et~al.}{1996}]{Christensen-Dalsgaard1996}
Christensen-Dalsgaard, J., D{\"{a}}ppen, W., Ajukov, S.~V., Anderson, E.~R.,
  Antia, H.~M., Basu, S., Baturin, V.~A., Berthomieu, G., Chaboyer, B., Chitre,
  S.~M., Cox, A.~N., Demarque, P., Donatowicz, J., Dziembowski, W.~A., Gabriel,
  M., Gough, D.~O., Guenther, D.~B., Guzik, J.~A., Harvey, J.~W., Hill, F.,
  Houdek, G., Iglesias, C.~A., Kosovichev, A.~G., Leibacher, J.~W., Morel, P.,
  Proffitt, C.~R., Provost, J., Reiter, J., Rhodes, E.~J., Rogers, F.~J.,
  Roxburgh, I.~W., Thompson, M.~J., and Ulrich, R.~K.: 1996,
\newblock {\em Science} {\bf 272(5266)}, 1286,
\newblock
  {\small[\href{http://science.sciencemag.org/content/272/5266/1286/tab-article-info}{URL}]}

\bibitem[\protect\astroncite{Dahlen and Tromp}{1998}]{Dahlen1998}
Dahlen, F.~A. and Tromp, J.: 1998,
\newblock {\em {Theoretical Global Seismology}},
\newblock Princeton University Press, Princeton, New Jersey, first edit edition

\bibitem[\protect\astroncite{Dikpati and Gilman}{2009}]{Dikpati2009}
Dikpati, M. and Gilman, P.~A.: 2009,
\newblock {\em Space Science Reviews} {\bf 144(1-4)}, 67,
\newblock
  {\small[\href{http://link.springer.com/10.1007/978-1-4419-0239-9{\_}5}{URL}]},
\newblock {\small[\href{http://dx.doi.org/10.1007/978-1-4419-0239-9_5}{DOI}]}

\bibitem[\protect\astroncite{Duez et~al.}{2008}]{Duez2008}
Duez, V., Mathis, S., Brun, A.~S., and Turck-Chi{\`{e}}ze, S.: 2008,
\newblock in K.~G. Strassmeier, A.~G. Kosovichev, and J.~E. Beckman (eds.),
  {\em Cosmic Magnetic Fields: From Planets, to Stars and Galaxies. Proceedings
  of the International Astronomical Union.}, No. S259, pp 177--184, Cambridge
  University Press,
\newblock
  {\small[\href{http://adsabs.harvard.edu/abs/2009IAUS..259..177D}{URL}]},
\newblock {\small[\href{http://dx.doi.org/10.1017/S1743921309030427}{DOI}]}

\bibitem[\protect\astroncite{Duez et~al.}{2010}]{Duez2010}
Duez, V., Mathis, S., and Turck-Chi{\`{e}}ze, S.: 2010,
\newblock {\em Monthly Notices of the Royal Astronomical Society} {\bf 402(1)},
  271,
\newblock
  {\small[\href{http://adsabs.harvard.edu/abs/2010MNRAS.402..271D}{URL}]},
\newblock
  {\small[\href{http://dx.doi.org/10.1111/j.1365-2966.2009.15955.x}{DOI}]}

\bibitem[\protect\astroncite{Edmonds}{1960}]{Edmonds}
Edmonds, A.~R.: 1960,
\newblock {\em {Angular momentum in quantum mechanics}},
\newblock Princeton University Press, Princeton, New Jersey, second edition

\bibitem[\protect\astroncite{Fan}{2009}]{Fan2009}
Fan, Y.: 2009,
\newblock {\em Living Reviews in Solar Physics} {\bf 6(1)}, 4,
\newblock {\small[\href{http://link.springer.com/10.12942/lrsp-2009-4}{URL}]},
\newblock {\small[\href{http://dx.doi.org/10.12942/lrsp-2009-4}{DOI}]}

\bibitem[\protect\astroncite{Garc{\'{i}}a et~al.}{2010}]{Garcia2010}
Garc{\'{i}}a, R.~A., Mathur, S., Salabert, D., Ballot, J., R{\'{e}}gulo, C.,
  Metcalfe, T.~S., and Baglin, A.: 2010,
\newblock {\em Science} {\bf 329(5995)}, 1032,
\newblock
  {\small[\href{http://adsabs.harvard.edu/abs/2010Sci...329.1032G}{URL}]},
\newblock {\small[\href{http://dx.doi.org/10.1126/science.1191064}{DOI}]}

\bibitem[\protect\astroncite{Giles et~al.}{1997}]{Giles1997}
Giles, P.~M., Duvall, T.~L., Scherrer, P.~H., and Bogart, R.~S.: 1997,
\newblock {\em Nature} {\bf 390(6655)}, 52,
\newblock
  {\small[\href{http://adsabs.harvard.edu/abs/1997Natur.390...52G}{URL}]},
\newblock {\small[\href{http://dx.doi.org/10.1038/36294}{DOI}]}

\bibitem[\protect\astroncite{Gough and Thompson}{1990}]{Gough1990}
Gough, D.~O. and Thompson, M.~J.: 1990,
\newblock {\em Monthly Notices of the Royal Astronomical Society} {\bf 242(1)},
  25,
\newblock
  {\small[\href{http://mnras.oxfordjournals.org/cgi/doi/10.1093/mnras/242.1.25}{URL}]},
\newblock {\small[\href{http://dx.doi.org/10.1093/mnras/242.1.25}{DOI}]}

\bibitem[\protect\astroncite{Haber et~al.}{2002}]{Haber2002}
Haber, D.~A., Hindman, B.~W., Toomre, J., Bogart, R.~S., Larsen, R.~M., and
  Hill, F.: 2002,
\newblock {\em The Astrophysical Journal} {\bf 570(2)}, 855,
\newblock
  {\small[\href{http://adsabs.harvard.edu/abs/2002ApJ...570..855H}{URL}]},
\newblock {\small[\href{http://dx.doi.org/10.1086/339631}{DOI}]}

\bibitem[\protect\astroncite{Hanasoge}{2017}]{Hanasoge2017}
Hanasoge, S.~M.: 2017,
\newblock {\em Monthly Notices of the Royal Astronomical Society} {\bf 470(3)},
  2780,
\newblock
  {\small[\href{http://adsabs.harvard.edu/abs/2017MNRAS.470.2780H}{URL}]},
\newblock {\small[\href{http://dx.doi.org/10.1093/mnras/stx1342}{DOI}]}

\bibitem[\protect\astroncite{Hekker and
  Christensen-Dalsgaard}{2017}]{Hekker2016}
Hekker, S. and Christensen-Dalsgaard, J.: 2017,
\newblock {\em The Astronomy and Astrophysics Review} {\bf 25}, 1,
\newblock {\small[\href{http://arxiv.org/abs/1609.07487}{URL}]}

\bibitem[\protect\astroncite{Herzberg}{2016}]{Herzberg2016}
Herzberg, W.: 2016,
\newblock {\em Ph.D. thesis}, Albert-Ludwigs Universit{\"{a}}t Freiburg,
\newblock {\small[\href{http://dx.doi.org/10.6094/UNIFR/11692}{DOI}]}

\bibitem[\protect\astroncite{Howe}{2009}]{Howe2009}
Howe, R.: 2009,
\newblock {\em Living Reviews in Solar Physics} {\bf 6(1)}, 1,
\newblock {\small[\href{http://link.springer.com/10.12942/lrsp-2009-1}{URL}]},
\newblock {\small[\href{http://dx.doi.org/10.12942/lrsp-2009-1}{DOI}]}

\bibitem[\protect\astroncite{Jefferies et~al.}{1990}]{Jefferies1990}
Jefferies, S.~M., Duvall, T.~L., Harvey, J.~W., and Pomerantz, M.~A.: 1990,
\newblock in Y. Osaki and H. Shibahashi (eds.), {\em Progress of Seismology of
  the Sun and Stars}, Vol. 367, pp 135--143, Springer Berlin Heidelberg,
  Berlin, Heidelberg

\bibitem[\protect\astroncite{Jimenez-Reyes et~al.}{1998}]{Jimenez-Reyes1998}
Jimenez-Reyes, S.~J., Regulo, C., Palle, P.~L., and {Roca Cortes}, T.: 1998,
\newblock {\em Astronomy and Astrophysics} {\bf 329}, 1119,
\newblock
  {\small[\href{http://adsabs.harvard.edu/abs/1998A{\%}26A...329.1119J}{URL}]}

\bibitem[\protect\astroncite{Karak et~al.}{2014}]{Karak2014}
Karak, B.~B., Jiang, J., Miesch, M.~S., Charbonneau, P., and Choudhuri, A.~R.:
  2014,
\newblock {\em Space Science Reviews} {\bf 186(1)}, 561,
\newblock
  {\small[\href{http://adsabs.harvard.edu/abs/2014SSRv..186..561K}{URL}]},
\newblock {\small[\href{http://dx.doi.org/10.1007/s11214-014-0099-6}{DOI}]}

\bibitem[\protect\astroncite{Kiefer et~al.}{2017}]{Kiefer2017}
Kiefer, R., Schad, A., Davies, G., and Roth, M.: 2017,
\newblock {\em Astronomy {\&} Astrophysics} {\bf 598}, A77,
\newblock {\small[\href{http://dx.doi.org/10.1051/0004-6361/201628469}{DOI}]}

\bibitem[\protect\astroncite{Komm et~al.}{2000}]{Komm2000a}
Komm, R.~W., Howe, R., and Hill, F.: 2000,
\newblock {\em The Astrophysical Journal} {\bf 531(2)}, 1094,
\newblock
  {\small[\href{http://adsabs.harvard.edu/abs/2000ApJ...531.1094K}{URL}]},
\newblock {\small[\href{http://dx.doi.org/10.1086/308518}{DOI}]}

\bibitem[\protect\astroncite{Lavely and Ritzwoller}{1992}]{Lavely1992}
Lavely, E.~M. and Ritzwoller, M.~H.: 1992,
\newblock {\em Philosophical Transactions: Physical Sciences and Engineering}
  {\bf 339(1655)}, 431

\bibitem[\protect\astroncite{Libbrecht and Woodard}{1990}]{Libbrecht1990}
Libbrecht, K.~G. and Woodard, M.~F.: 1990,
\newblock {\em Nature} {\bf 345(6278)}, 779,
\newblock
  {\small[\href{http://adsabs.harvard.edu/abs/1990Natur.345..779L}{URL}]},
\newblock {\small[\href{http://dx.doi.org/10.1038/345779a0}{DOI}]}

\bibitem[\protect\astroncite{Lynden-Bell and Ostriker}{1967}]{Lynden-Bell1967}
Lynden-Bell, D. and Ostriker, J. P.~J.: 1967,
\newblock {\em Monthly Notices of the Royal Astronomical Society} {\bf 136(3)},
  293,
\newblock
  {\small[\href{http://adsabs.harvard.edu/abs/1967MNRAS.136..293L}{URL}]},
\newblock {\small[\href{http://dx.doi.org/10.1093/mnras/136.3.293}{DOI}]}

\bibitem[\protect\astroncite{Mathis and Zahn}{2005}]{Mathis2005}
Mathis, S. and Zahn, J.-P.: 2005,
\newblock {\em Astronomy and Astrophysics} {\bf 440(2)}, 653,
\newblock
  {\small[\href{http://adsabs.harvard.edu/abs/2005A{\%}26A...440..653M}{URL}]},
\newblock {\small[\href{http://dx.doi.org/10.1051/0004-6361:20052640}{DOI}]}

\bibitem[\protect\astroncite{Mestel and Moss}{1977}]{Mestel1977}
Mestel, L. and Moss, D.~L.: 1977,
\newblock {\em Monthly Notices of the Royal Astronomical Society} {\bf 178(1)},
  27,
\newblock
  {\small[\href{http://adsabs.harvard.edu/abs/1977MNRAS.178...27M}{URL}]},
\newblock {\small[\href{http://dx.doi.org/10.1093/mnras/178.1.27}{DOI}]}

\bibitem[\protect\astroncite{Miesch and Teweldebirhan}{2016}]{Miesch2016}
Miesch, M. and Teweldebirhan, K.: 2016,
\newblock {\em Advances in Space Research} {\bf 58(8)}, 1571,
\newblock {\small[\href{http://dx.doi.org/10.1016/j.asr.2016.02.018}{DOI}]}

\bibitem[\protect\astroncite{Nielsen et~al.}{2015}]{Nielsen2015}
Nielsen, M.~B., Schunker, H., Gizon, L., and Ball, W.~H.: 2015,
\newblock {\em Astronomy {\&} Astrophysics} {\bf 582}, A10,
\newblock
  {\small[\href{http://adsabs.harvard.edu/abs/2015A{\%}26A...582A..10N}{URL}]},
\newblock {\small[\href{http://dx.doi.org/10.1051/0004-6361/201526615}{DOI}]}

\bibitem[\protect\astroncite{Phinney and Burridge}{1973}]{Phinney1973}
Phinney, R.~A. and Burridge, R.: 1973,
\newblock {\em Geophysical Journal International} {\bf 34(4)}, 451,
\newblock
  {\small[\href{http://onlinelibrary.wiley.com/doi/10.1111/j.1365-246X.1973.tb02407.x/abstract}{URL}]},
\newblock
  {\small[\href{http://dx.doi.org/10.1111/j.1365-246X.1973.tb02407.x}{DOI}]}

\bibitem[\protect\astroncite{Racah}{1942}]{Racah1942}
Racah, G.: 1942,
\newblock {\em Physical Review} {\bf 62(9-10)}, 438,
\newblock
  {\small[\href{http://journals.aps.org/pr/abstract/10.1103/PhysRev.62.438}{URL}]},
\newblock {\small[\href{http://dx.doi.org/10.1103/PhysRev.62.438}{DOI}]}

\bibitem[\protect\astroncite{Regge}{1958}]{Regge1958}
Regge, T.: 1958,
\newblock {\em Il Nuovo Cimento} {\bf 10(3)}, 544,
\newblock {\small[\href{http://link.springer.com/10.1007/BF02859841}{URL}]},
\newblock {\small[\href{http://dx.doi.org/10.1007/BF02859841}{DOI}]}

\bibitem[\protect\astroncite{Ritzwoller and Lavely}{1991}]{Ritzwoller1991a}
Ritzwoller, M.~H. and Lavely, E.~M.: 1991,
\newblock {\em The Astrophysical Journal} {\bf 369}, 557,
\newblock {\small[\href{http://adsabs.harvard.edu/doi/10.1086/169785}{URL}]},
\newblock {\small[\href{http://dx.doi.org/10.1086/169785}{DOI}]}

\bibitem[\protect\astroncite{Roth}{2002}]{Roth2002}
Roth, M.: 2002,
\newblock {\em Ph.D. thesis}, Albert-Ludwigs-Universit{\"{a}}t Freiburg,
\newblock {\small[\href{https://freidok.uni-freiburg.de/data/512}{URL}]},
\newblock {\small[\href{http://dx.doi.org/URN:
  urn:nbn:de:bsz:25-opus-5126}{DOI}]}

\bibitem[\protect\astroncite{Roth and Stix}{2008}]{Roth2008}
Roth, M. and Stix, M.: 2008,
\newblock {\em Solar Physics} {\bf 251(1-2)}, 77,
\newblock {\small[\href{http://dx.doi.org/10.1007/s11207-008-9232-6}{DOI}]}

\bibitem[\protect\astroncite{Salabert et~al.}{2016}]{Salabert2016a}
Salabert, D., R{\'{e}}gulo, C., Garc{\'{i}}a, R.~A., Beck, P.~G., Ballot, J.,
  Creevey, O.~L., {P{\'{e}}rez Hern{\'{a}}ndez}, F., {do Nascimento Jr.},
  J.-D., Corsaro, E., Egeland, R., Mathur, S., Metcalfe, T.~S., Bigot, L.,
  Ceillier, T., and Pall{\'{e}}, P.~L.: 2016,
\newblock {\em Astronomy {\&} Astrophysics} {\bf 589}, A118,
\newblock
  {\small[\href{http://adsabs.harvard.edu/abs/2016A{\%}26A...589A.118S}{URL}]},
\newblock {\small[\href{http://dx.doi.org/10.1051/0004-6361/201527978}{DOI}]}

\bibitem[\protect\astroncite{Schad}{2013}]{Schad2013}
Schad, A.: 2013,
\newblock {\em Ph.D. thesis}, Albert-Ludwigs-Universit{\"{a}}t Freiburg

\bibitem[\protect\astroncite{Schad et~al.}{2011}]{Schad2011}
Schad, A., Timmer, J., and Roth, M.: 2011,
\newblock {\em The Astrophysical Journal} {\bf 734(2)}, 97,
\newblock
  {\small[\href{http://stacks.iop.org/0004-637X/734/i=2/a=97?key=crossref.dadfa4c048560d4a4061392ba3bf8e06}{URL}]},
\newblock {\small[\href{http://dx.doi.org/10.1088/0004-637X/734/2/97}{DOI}]}

\bibitem[\protect\astroncite{Schad et~al.}{2013}]{Schad2013a}
Schad, A., Timmer, J., and Roth, M.: 2013,
\newblock {\em The Astrophysical Journal} {\bf 778(2)}, L38,
\newblock
  {\small[\href{http://stacks.iop.org/2041-8205/778/i=2/a=L38?key=crossref.b8638d3f555cf0c0fc75f57e5ec12f8f}{URL}]},
\newblock {\small[\href{http://dx.doi.org/10.1088/2041-8205/778/2/L38}{DOI}]}

\bibitem[\protect\astroncite{Schou et~al.}{1998}]{Schou1998}
Schou, J., Antia, H.~M., Basu, S., Bogart, R.~S., Bush, R.~I., Chitre, S.~M.,
  Christensen‐Dalsgaard, J., {Di Mauro}, M.~P., Dziembowski, W.~A.,
  Eff‐Darwich, A., Gough, D.~O., Haber, D.~A., Hoeksema, J.~T., Howe, R.,
  Korzennik, S.~G., Kosovichev, A.~G., Larsen, R.~M., Pijpers, F.~P., Scherrer,
  P.~H., Sekii, T., Tarbell, T.~D., Title, A.~M., Thompson, M.~J., and Toomre,
  J.: 1998,
\newblock {\em The Astrophysical Journal} {\bf 505(1)}, 390,
\newblock
  {\small[\href{http://adsabs.harvard.edu/abs/1998ApJ...505..390S}{URL}]},
\newblock {\small[\href{http://dx.doi.org/10.1086/306146}{DOI}]}

\bibitem[\protect\astroncite{Unno et~al.}{1989}]{Unno1989}
Unno, W., Osaki, Y., Ando, H., Saio, H., and Shibahashi, H.: 1989,
\newblock {\em {Nonradial oscillations of stars}},
\newblock University of Tokyo Press, Tokyo, second edition

\bibitem[\protect\astroncite{Woodard and Noyes}{1985}]{Woodard1985}
Woodard, M.~F. and Noyes, R.~W.: 1985,
\newblock {\em Nature} {\bf 318(6045)}, 449,
\newblock
  {\small[\href{http://adsabs.harvard.edu/abs/1985Natur.318..449W}{URL}]},
\newblock {\small[\href{http://dx.doi.org/10.1038/318449a0}{DOI}]}

\bibitem[\protect\astroncite{Zhao et~al.}{2013}]{Zhao2013a}
Zhao, J., Bogart, R.~S., Kosovichev, A.~G., Duvall, T.~L., and Hartlep, T.:
  2013,
\newblock {\em The Astrophysical Journal Letters} {\bf 774(2)}, L29,
\newblock
  {\small[\href{http://adsabs.harvard.edu/abs/2013ApJ...774L..29Z}{URL}]},
\newblock {\small[\href{http://dx.doi.org/10.1088/2041-8205/774/2/L29}{DOI}]}

\end{thebibliography}
\end{document}